\documentclass[a4paper,fleqn]{style_neural_networks/cas-dc}

\usepackage{listings}
\usepackage{latexsym}
\usepackage{amsthm}
\usepackage{longtable}
\usepackage{chemarr}
\usepackage{rotating}
\usepackage[english]{babel}
\usepackage{bm}
\usepackage{microtype}
\usepackage{csquotes}
\usepackage{soul}
\usepackage{placeins}
\usepackage[authoryear,longnamesfirst]{natbib}
\bibpunct{(}{)}{;}{a}{,}{,}
\let\parencite\citep

\begin{document}

\let\WriteBookmarks\relax
\def\floatpagepagefraction{1}
\def\textpagefraction{.001}

\shorttitle{Hyperparameter robustness in reservoirs}
\shortauthors{Churchland}

\title [mode = title]{Determinants of hyperparameter robustness in connectome reservoir computing}

\author[1]{Miles Walter Churchland}
\credit{Conceptualization, methodology, software, formal analysis, visualization, writing - original draft, writing - review and editing}
\author[2]{Raul de Palma Aristides}
\credit{Review, assistance with plotting}
\author[1]{Jordi Garcia-Ojalvo}
\credit{Review}
\author[3]{Anna Ritz}
\credit{Review, Codebase improvements }
\author[4]{Greg Anderson}
\credit{Review}
\author[2]{Miguel C. Soriano}
\credit{Review}

\affiliation[1]{organization={Department of Medicine and Life Sciences, Universitat Pompeu Fabra},
                city={Barcelona},
                country={Spain}}

\affiliation[2]{organization={Institute for Cross-Disciplinary Physics and Complex Systems (IFISC), CSIC--UIB},
                city={Palma},
                state={Mallorca},
                country={Spain}}

\affiliation[3]{organization={Biology Department, Reed College},
                city={Portland},
                state={OR},
                country={United States}}

\affiliation[4]{organization={Department of Computer Science, Reed College},
                city={Portland},
                state={OR},
                country={United States}}
\begin{abstract}
Reservoir computing provides a controlled setting for studying how recurrent network architecture shapes computation: input signals are projected into a high-dimensional state space by a fixed nonlinear dynamical system, and only the readout is trained. However, reservoir performance can be strongly dependent on hyperparameters; this paper asks which recurrent network features support robustness to those parameter changes. We characterize computational performance using memory capacity (MC), truncated single-delay information-processing capacity (IPC), and kernel rank (KR). Generalization across input histories is measured using generalization rank (GR), while hyperparameter robustness is quantified using the coefficient of variation (CV) of each metric across sweeps of target spectral radius, input scaling, leak rate, and neuron bias.

To examine the architectural determinants of robustness, we construct controlled perturbations that alter connectivity topology, excitatory/inhibitory sign structure, weight magnitudes, and weight placement while preserving complementary properties. Across these experiments, the \textit{C.\ elegans} connectome consistently occupies a relatively low-variance regime.

The central result is a performance--robustness tradeoff: architecture variants with higher task-agnostic performance also tend to exhibit greater hyperparameter sensitivity and poorer common-tail generalization. Across the E/I edge balance sweeps and shuffle controls, this tradeoff is closely associated with the raw spectral radius before normalization. Because every perturbed matrix is subsequently rescaled to the same target radius, matrices with lower raw spectral radius receive greater global amplification of their recurrent weights. The observed differences among architectures therefore characterize the joint effects of structural variation and architecture-specific global rescaling under spectral-radius normalization.

\end{abstract}

\begin{highlights}
\item Connectome-derived reservoir architectures can occupy low hyperparameter-sensitivity regimes.
\item Higher task-agnostic performance coincides with greater hyperparameter sensitivity and poorer common-tail generalization.
\item Raw spectral radius determines recurrent-weight scaling under standard spectral-radius normalization.
\end{highlights}


\maketitle

\section{Introduction}
Reservoir computing uses a fixed recurrent dynamical system to project input sequences into a high-dimensional state space, after which only a readout is trained \parencite{jaeger_echo_state_2001,maass_lsm_2002}. This separation between fixed recurrent weights and trained readouts makes reservoir models useful for studying how recurrent neural network architecture shapes computation. At the same time, reservoir dynamics and performance can vary strongly with hyperparameters such as target spectral radius, input scaling, leak rate, and neuronal bias \parencite{Lukosevicius2012PracticalESN,Matzner2022HyperparameterTuningESN}.

Random recurrent networks are commonly used as reservoir baselines because they can generate diverse nonlinear features and support memory and separability across suitable dynamical regimes \parencite{Lukosevicius2012PracticalESN}. Biological connectomes provide a contrasting class of architectures whose wiring is non-random, sparse, directed, weighted, and constrained by biological function \parencite{gabalda2018}. The \textit{C.\ elegans} connectome is particularly useful because it is compact, well characterized, and experimentally grounded. This contrast motivates asking whether structured biological connectivity can support reservoir behavior that is not only computationally effective, but also stable across hyperparameter variation.

Biologically inspired networks have also been used directly as reservoirs or reservoir-like substrates. Studies have treated biological connectivity as a fixed graph: \cite{KAWAI201915} compared echo-state networks with small-world structure to models based on human cortical connectivity, while \cite{morra_daley_2022} replaced the reservoir layer of an echo state network (ESN) with fruit-fly connectome-derived topologies. Later work using the fruit-fly lateral horn found that connectome-based reservoirs supported multifunctionality across a broader region of hyperparameter-space than Erd\H{o}s--R\'enyi controls \parencite{morra_fly_rc_2023}. Other work has moved from anatomical graphs to living substrates, using cultured cortical neurons as the reservoir itself \parencite{iannello_brc_2025}. 

We use an anatomical \textit{C.\ elegans} chemical-synapse neural connectome from EleganSign \parencite{fenyves_szilágyi_zsolt_vassy_csaba_2020,elegansign_web} as a structured recurrent weight matrix, then apply controlled perturbations that vary topology, inferred excitatory/inhibitory sign structure, weight distribution, weight placement, and sign placement.\footnote{Throughout this paper, E/I edge balance refers to the ratio or fraction of excitatory to inhibitory synaptic connections in the modeled reservoir; it is an edge-level quantity, not a count of excitatory or inhibitory neurons.} These reservoirs are evaluated with task-agnostic metrics--hyperparameter robustness is quantified by the degree to which these metrics change across our grid of hyperparameters.

A key interpretive issue arises from the standard spectral-radius normalization used in these comparisons. Each structural perturbation first changes the unnormalized recurrent matrix and potentially its raw spectral radius. Rescaling every matrix to a common target radius then applies a perturbation-specific global weight factor. Raw spectral radius is therefore not merely a measured property of each architecture: it also determines how strongly that architecture's recurrent weights are amplified or attenuated before evaluation.

The results identify a tradeoff between performance and hyperparameter robustness. Architecture variants with higher performance show both greater sensitivity to hyperparameter choice and poorer common-tail generalization. These patterns are closely associated with raw spectral radius and the resulting normalization factor.

\section{Methods}
\label{sec:methods}
\subsection{Reservoir model and hyperparameters}
We use echo state networks, a discrete-time reservoir-computing model in which recurrent weights are fixed and only linear readouts are trained using linear regression \parencite{jaeger_echo_state_2001,Lukosevicius2012PracticalESN,sun_systematic_review_2024}. The state update rule is
\begin{align}
{x}_{t+1} = &(1-\alpha)\,x_t \nonumber \\
& +\alpha\,\tanh\!\left(W^{\mathrm{res}} x_t + s_{\mathrm{in}}W^{\mathrm{in}} u_{t} + b \right),
\end{align}
where \(t\) is the discrete time index, the vector \(x_t\in\mathbb{R}^{N}\) is the reservoir state at time \(t\), and \(u_t\in\mathbb{R}^{m}\) is the input vector. \(N\) is the number of reservoir units (neurons) and $m$ is the number of inputs. 
The output \(y_t\in\mathbb{R}^{p}\) is given by:
\begin{align}
y_t &= W^{\mathrm{out}}\,x_t.
\end{align}
The matrices \(W^{\mathrm{res}}\in\mathbb{R}^{N\times N}\), \(W^{\mathrm{in}}\in\mathbb{R}^{N\times m}\), and \(W^{\mathrm{out}}\in\mathbb{R}^{p\times N}\) are the recurrent, input, and readout weight matrices, respectively.
The experiments use scalar inputs, so \(m=1\). 
The scalar \(s_{\mathrm{in}}\) sets the input scale, \(b\in\mathbb{R}^{N}\) is a fixed per-neuron bias vector, and \(\alpha\in(0,1]\) is the leak rate. The \(\tanh\) activation is applied elementwise.\footnote{The readout dimension is \(p=5\) for the truncated information processing capacity (IPC) diagnostic, defined in Section~\ref{sec:ipc}, so that multiple polynomial orders can be solved in a single batch; otherwise \(p=1\).} Only \(W^{\mathrm{out}}\) is trained, and before evaluation each instantiated reservoir is rescaled to the selected target spectral radius following standard ESN spectral-radius normalization \citep{Lukosevicius2012PracticalESN}.
For each trial, the entries of the base input-weight vector
\(W^{\mathrm{in}}\in\mathbb{R}^{N\times 1}\) are sampled independently as
\[
W^{\mathrm{in}}_{i}\sim\mathcal{N}(0,1),
\qquad i=1,\ldots,N.
\]
The scalar input is projected to all \(N=297\) neurons through
\(s_{\mathrm{in}}W^{\mathrm{in}}\). The same realization of
\(W^{\mathrm{in}}\) is used across the complete hyperparameter grid within
each trial; only \(s_{\mathrm{in}}\) changes. A new realization is sampled
for each trial.

\label{sec:hyperparameters}
The experiments sweep four hyperparameters: target spectral radius, leak rate, input scaling, and neuron-bias. The spectral radius \(\rho(W^{\mathrm{res}})\) sets the strength of recurrent feedback, where
\[
\rho(W) = \max_{1 \le i \le N} |\lambda_i|.
\]
Here \(W\in\mathbb{R}^{N\times N}\) is a weight matrix and \(\lambda_i\) is its \(i\)th eigenvalue. In reservoir computing, recurrent matrices are commonly rescaled to a target spectral radius near one to keep the dynamics in a useful operating range \parencite{jaeger_echo_state_2001,Lukosevicius2012PracticalESN}, although reservoirs can remain computationally useful at values below or above this range \parencite{gallicchio_chasing_2019}. In our experiments, network perturbations are applied first, producing an unnormalized recurrent matrix \(W'\). We define its spectral radius before normalization as the raw spectral radius,
\[
\rho_{\mathrm{raw}}
\equiv \rho(W')
=\max_i |\lambda_i(W')|.
\]
We then rescale this matrix as
\[
W^{\mathrm{res}}
=
\frac{\rho_{\mathrm{target}}}{\rho_{\mathrm{raw}}}W',
\qquad
\rho(W^{\mathrm{res}})=\rho_{\mathrm{target}}.
\]
Here \(\lambda_i(W')\) is the \(i\)th eigenvalue of the perturbed matrix, \(\rho_{\mathrm{target}}\) is the selected target spectral radius, and \(W^{\mathrm{res}}\) is the normalized recurrent matrix used in the state update. Because the normalization factor is \(\rho_{\mathrm{target}}/\rho_{\mathrm{raw}}\), any perturbation that changes the raw spectral radius also changes the global scaling applied to the recurrent weights. This procedure holds the achieved spectral radius fixed across architectures, but it does not hold the normalization factor or the weight scale fixed. Conversely, applying a fixed scaling factor would allow the achieved spectral radius to vary across architectures. For matrices with different \(\rho_{\mathrm{raw}}\), scalar rescaling alone cannot hold both the achieved spectral radius and the rescaling factor constant. The experiments therefore show how the perturbed architectures behave after spectral-radius normalization, rather than estimating structural effects isolated from their raw-spectral-radius consequences.

For the remaining hyperparameters, the leak rate \(\alpha\) sets the
effective timescale of the reservoir, and input scaling
\(s_{\mathrm{in}}\) sets the strength of the external drive. The
neuron-bias parameter \(\beta\) determines the maximum absolute bias.
When \(\beta=0\), all neuron biases are zero. When \(\beta=0.1\), the
biases are sampled independently as
\[
b_i\sim\mathcal{U}(-0.1,0.1),
\qquad i=1,\ldots,N.
\]
One bias vector is sampled for each trial and held fixed across all 48
hyperparameter combinations for which \(\beta=0.1\); it is not redrawn
when the spectral radius, leak rate, or input scaling changes. A new
bias vector is sampled for each trial.
The sweep values were chosen as a compact sensitivity grid. Every architecture is evaluated on the same grid. The target spectral-radius values \(\{0.6,0.8,0.95,1.05\}\) bracket the standard near-one operating regime while including more contractive and slightly supercritical targets \parencite{Lukosevicius2012PracticalESN,gallicchio_chasing_2019}. The leak values \(\{0.6,0.8,1.0\}\) sample moderate leaky integration through the non-leaky update, the input-scaling values \(\{0.1,0.5,1.0,1.5\}\) span weak to strong input drive, and the bias values \(\{0.0,0.1\}\) compare an unbiased reservoir with a small fixed per-neuron offset. These choices cover standard ESN design knobs, including bias terms that can affect reservoir performance \parencite{Lukosevicius2012PracticalESN,platt_systematic_2022,Matzner2022HyperparameterTuningESN}. Their Cartesian product contains \(4\times3\times4\times2=96\) hyperparameter combinations. For each trial and task-agnostic metric, these combinations produce 96 metric values. We average those values to obtain the mean metric value and summarize their relative dispersion with the coefficient of variation (CV), defined in Section~\ref{sec:hyperparameter_invariance}. Higher values indicate greater performance for IPC, KR, and MC, whereas lower GR indicates better common-tail generalization. Thus, each trial produces one mean metric value and one CV value for each metric.

\subsection{Connectome data and preprocessing}
\label{sec:connectome_dataset}
We use an anatomical \textit{C.\ elegans} chemical-synapse neural connectome with predicted chemical polarity labels derived from EleganSign \parencite{fenyves_szilágyi_zsolt_vassy_csaba_2020}. Here, a polarity label specifies whether a chemical connection is predicted to have an excitatory (positive) or inhibitory (negative) effect. In our reservoir model, each of the 297 neurons in the connectome is represented as one reservoir unit, giving a total reservoir size of \(N=297\), and the directed chemical synapses define the recurrent matrix \(W^{\mathrm{res}}\). A chemical connection from neuron \(i\) to neuron \(j\) adds a directed edge with magnitude equal to the number of synaptic contacts from \(i\) to neuron \(j\) \parencite{Varshney2011,Cook2019,Casal2020,churchland_garcia-ojalvo_2026}. Self-loops, for which \(i=j\), are removed during preprocessing. For chemical edges with a predicted polarity label, we multiply this synaptic contact-count magnitude by the inferred sign, so predicted excitatory chemical synapses enter the matrix with positive weights and predicted inhibitory chemical synapses enter with negative weights. Unpredicted/Complex chemical edges are handled by the preprocessing approaches described below. The resulting signed, weighted adjacency matrix is the fixed recurrent reservoir used in the reservoir update above: inputs drive the reservoir through \(W^{\mathrm{in}}\), recurrent activity propagates through the connectome-derived \(W^{\mathrm{res}}\), and only the output readout \(W^{\mathrm{out}}\) is trained. Before evaluation, each reservoir instance is rescaled to the target spectral radius described in Sec.~\ref{sec:hyperparameters}.

Polarity labels apply only to chemical synapses. We follow the EleganSign chemical-edge export as the source database for the recurrent matrix \parencite{fenyves_szilágyi_zsolt_vassy_csaba_2020}. EleganSign predicts excitatory or inhibitory effects for chemical synapses, so its database does not include electrical synapses. The labels are also predictions rather than direct functional measurements, because they are inferred from gene-expression and neurotransmitter logic rather than from voltage- or calcium-response measurements \parencite{fenyves_szilágyi_zsolt_vassy_csaba_2020}. EleganSign predicts signs for 1740 synapses, while 1864 synapses have unknown or complex sign predictions.  For each repeat, these 1864 edges are randomly permuted, exactly 452 are assigned negative signs, and the remaining 1412 are assigned positive signs. This fixed count matches the negative ratio for connections which EleganSign provides a polarity prediction, \(p_-\approx0.24\). To make the analysis less dependent on a single treatment of these 1864 edges, selected experiments are repeated on a predicted-polarity-only version, where unknown or complex-sign connections are removed (Supplemental Figure~\ref{fig:removed_frac}).

\subsection{Task-agnostic reservoir metrics}
We measure four task-agnostic reservoir metrics: kernel rank (KR), generalization rank (GR), memory capacity (MC), and a truncated single-delay information-processing-capacity (IPC) diagnostic \parencite{maass_comp_power_2004,legenstein_maass_2007,busing_connectivity_2010,Jaeger2002STMESN,dambre_ipc_2012,pilati_ceni_michieletti_gallicchio_ricciardi_milano_2026}.

\subsubsection{Memory Capacity}
Memory capacity (MC) measures how well a linear readout can reconstruct past inputs from the current reservoir state. For each delay \(d\), a ridge-regression readout is trained to predict the delayed i.i.d. uniform input \(u(t-d)\) from \(x(t)\), and the squared Pearson correlation between predicted and true delayed input is recorded. Here, \(r_{\mathrm{score}}\) denotes this squared Pearson correlation, bounded between 0 and 1 \parencite{Jaeger2002STMESN}; its full definition is given in Supplemental Section~\ref{supp:r_score}. We approximate the total memory capacity by summing over delays up to \(D_{\max}=30\):
\[
\mathrm{MC}
=
\sum_{d=1}^{D_{\max}}
r_{\mathrm{score}}(u(t-d),\hat{u}_d(t)).
\]
The sensitivity of MC to this delay truncation is reported in Supplemental Section~\ref{supp:metric_truncation}.

\subsubsection{Truncated information-processing-capacity diagnostic}
\label{sec:ipc}
Information processing capacity (IPC) measures how well the reservoir can reconstruct nonlinear functions of delayed inputs \parencite{dambre_ipc_2012}. Here, IPC denotes a truncated single-delay diagnostic rather than the full IPC. Specifically, we evaluate univariate Legendre polynomial targets of one delayed i.i.d. uniform input at a time and omit pairwise and higher-order cross-delay product terms. This still measures the reservoir's ability to generate nonlinear transformations of input history, but it does not estimate the full capacity over all nonlinear input combinations. For Legendre polynomial orders \(K=\{1,2,3,4,5\}\) and delays \(d\in\{1,\ldots,D_{\max}\}\), with \(D_{\max}=30\), let \(z_{k,d}(t)=P_k(u(t-d))\). We compute
\[
\mathrm{IPC}
=
\sum_{d=1}^{D_{\max}}
\sum_{k\in K}
\max\!\left(0,\,
r_{\mathrm{score}}\!\left(z_{k,d}(t),\hat{z}_{k,d}(t)\right)
\right).
\]
Because cross-delay products are not included, the reported IPC values should be interpreted as an incomplete IPC-style statistic rather than an exhaustive estimate of full IPC. Sensitivity to the selected polynomial orders and maximum delay is reported in Supplemental Section~\ref{supp:metric_truncation}.

\subsubsection{Kernel Rank}
Kernel rank (KR) measures how many effectively independent reservoir activity dimensions are generated in response to distinct input histories \parencite{maass_comp_power_2004,legenstein_maass_2007,busing_connectivity_2010}. An activity dimension is an orthogonal direction in the \(N\)-dimensional reservoir state space along which the final states vary. A high KR is desirable because it indicates that the reservoir maps different inputs to diverse, linearly separable internal representations.

Following the reset, multi-stream protocol of \citet{vidamour_quantifying_2022}, we generate \(M=300\) independent input streams of length \(L=10\):
\[
u_t^{(m)}\sim\mathcal{U}(-1,1),
\qquad
m=1,\ldots,M,
\quad
t=1,\ldots,L.
\]
The reservoir is reset to the same zero initial state before every stream. We retain only the final state \(\mathbf{x}_L^{(m)}\in\mathbb{R}^{N}\) from each run and stack these states as rows of
\[
X_{\mathrm{KR}}
=
\begin{bmatrix}
(\mathbf{x}_L^{(1)})^\top\\
\vdots\\
(\mathbf{x}_L^{(M)})^\top
\end{bmatrix}
\in\mathbb{R}^{M\times N}.
\]
For any state-ensemble matrix \(X\), we define
\[
\begin{aligned}
\mathrm{KR}(X)
&:= E_{\mathrm{rank}}(\widetilde{X}), \\
\widetilde{X}
&= X-\mathbf{1}_M\bar{\mathbf{x}}(X)^\top, \\
\bar{\mathbf{x}}(X)
&= \frac{1}{M}X^\top\mathbf{1}_M.
\end{aligned}
\]
Thus, \(X\) is centered by subtracting each neuron's mean activity across streams. Let \(\sigma_1,\ldots,\sigma_r\) denote the nonzero singular values of \(\widetilde{X}\), and define
\[
p_i=\frac{\sigma_i}{\sum_{j=1}^{r}\sigma_j}.
\]
The effective rank is then
\[
E_{\mathrm{rank}}(\widetilde{X})
=
\exp\left(
-\sum_{i=1}^{r} p_i\log p_i
\right).
\]
We use the effective rank of \citet{roy_effective_rank_2007}, as applied in \parencite{love_task_agnostic_2021,Stepney2024PRCtutorial}. A high \(\mathrm{KR}(X_{\mathrm{KR}})\) is desirable and signifies that the reservoir separates different input histories into different final states.

\subsubsection{Generalization Rank}
Generalization rank (GR) measures how many effectively independent reservoir activity dimensions are attributable to differences in earlier input history \parencite{legenstein_maass_2007,busing_connectivity_2010}. Under the common-tail protocol, input streams have different earlier histories but share the same recent inputs. A low GR is desirable because variation caused by those earlier histories occupies fewer effective dimensions after the shared tail. Throughout this paper, better or poorer generalization refers specifically to this dependence on input history preceding the common tail.

We use the multi-stream, common-tail construction of \citet{vidamour_quantifying_2022}, which adapts the original rank-based generalization measure to uniformly distributed real-valued reservoir inputs and is summarized in the physical-reservoir tutorial of \citet{Stepney2024PRCtutorial}. As for KR, we generate \(M=300\) streams of length \(L=10\). The first \(L-L_c=7\) inputs are drawn independently for every stream,
\[
u_t^{(m)}\sim\mathcal{U}(-1,1),
\qquad
m=1,\ldots,M,
\quad
t=1,\ldots,7,
\]
whereas the final \(L_c=3\) inputs are sampled once and shared by all streams,
\[
u_t^{(m)}=v_t,
\qquad
t=8,\ldots,10,
\qquad
v_t\sim\mathcal{U}(-1,1).
\]
The reservoir is again reset to the same zero initial state before every stream, and the final states are stacked as
\[
X_{\mathrm{GR}}
=
\begin{bmatrix}
(\mathbf{x}_L^{(1)})^\top\\
\vdots\\
(\mathbf{x}_L^{(M)})^\top
\end{bmatrix}
\in\mathbb{R}^{M\times N}.
\]
Generalization rank is obtained by applying the same kernel-rank function to the common-tail state ensemble:
\[
\mathrm{GR}
:=
\mathrm{KR}\!\left(X_{\mathrm{GR}}\right).
\]

\begin{table*}[!t]
\centering
\scriptsize
\caption{Architecture families used in the reservoir perturbations.}
\label{tab:architecture_groups}
    \renewcommand{\arraystretch}{1.15}
    \setlength{\tabcolsep}{4pt}
\begin{tabular}{p{0.18\textwidth}p{0.35\textwidth}p{0.21\textwidth}p{0.22 \textwidth}}
\toprule
Architecture family & Manipulation & Preserved properties & Role \\
\midrule
Empirical baseline & \textit{C.\ elegans} topology, magnitudes, using EleganSign-predicted signs where available and assigning the sign of the remaining edges to match the observed sign ratios & All properties of preprocessing & Defines the reference regime \\
Sign-preserving weight changes & Replace weight magnitudes with signed unit, half-Gaussian, or uniform distributions & Connections, sign placement, and global sign ratio & Assesses whether the \textit{C.\ elegans}' performance--CV regime persists when signs are preserved but magnitudes are altered \\
Binary weights & Set all nonzero weights to one & Connections & Compares the signed, weighted reservoirs with a topology-only condition  \\
Connection Shuffles & Rewire edges using directed degree-preserving connection shuffles (Supplementary Section~\ref{supp:Connection-Shuffle}) & Node count, edge count, directed in- and out-degrees, signed-weight distribution & Examines behavior after changing local topology \\
Weight Shuffles & Shuffle signed weights across existing edges (Supplementary Section~\ref{supp:Weight-Shuffle}) & Topology, edge set, and signed-weight distribution & Examines behavior after changing local weight placement \\
Connection + Weight Shuffles & Rewire edges using directed connection shuffles, then shuffle weights across the rewired edges (Supplementary Sections~\ref{supp:Connection-Shuffle} and~\ref{supp:Weight-Shuffle}) & Directed in- and out-degrees, and signed-weight distribution & Examines behavior after changing both local topology and weight placement \\
Local Sign-placement Shuffles & Shuffle signs across existing weighted edges (Supplementary Section~\ref{supp:sign-pres_real_w}) & Topology, edge set, weight placement and weight-magnitude distribution & Examines behavior after changing local sign placement \\

Random-topology baselines & Use Gaussian-weight \textit{C.\ elegans}, Erd\H{o}s--R\'enyi, and Watts--Strogatz small-world topology controls \parencite{network_science,KAWAI201915,Wssmallworld} & Node count and the designated weight/sign regime & Compares behavior across broad topology classes \\
\bottomrule
\end{tabular}
\end{table*}
\begin{figure*}[htbp]
    \centering
    \includegraphics[width=\textwidth]{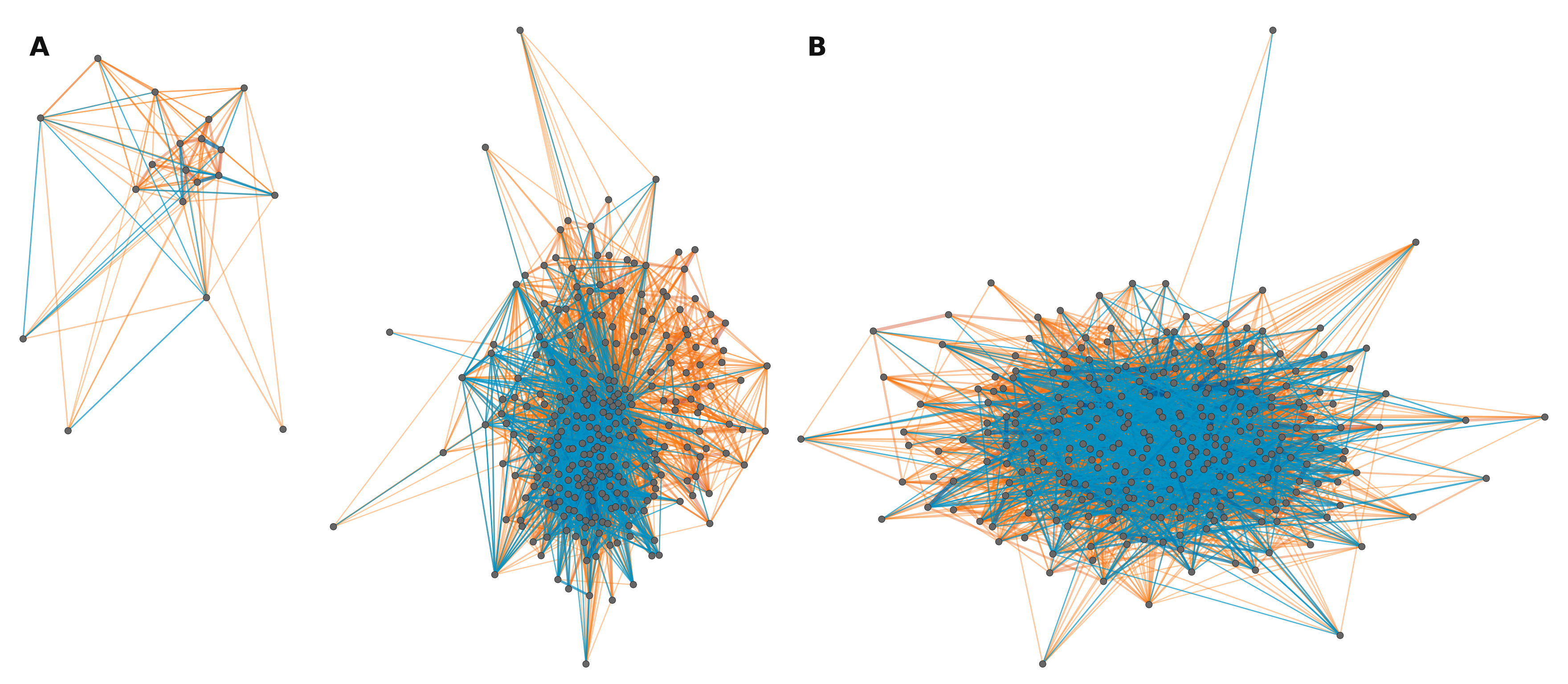}
    \caption{Representative network visualizations for the preprocessing and random-sign controls used in the architecture analyses. A: Empirical \textit{C.\ elegans} connectome, with predicted chemical polarity labels where available and unknown or complex-sign chemical edges assigned using fixed counts that match the E/I edge balance of the labeled connections. B: Connection-shuffled version of the network in A. All networks are directed, although edges are shown without directionality for visual clarity. Orange edges denote positive weights and blue edges negative weights, with color saturation indicating weight magnitude. The disconnected component visible in A corresponds to neurons of the pharyngeal nervous system. Following connection shuffling (B), this component becomes connected to the rest of the network, which may contribute to the resulting changes in task-agnostic metrics.}
    \label{fig:shuffle_four_panel}
\end{figure*}

\subsection{Hyperparameter-robustness metric}
\label{sec:hyperparameter_invariance}
To investigate stability under hyperparameter variation, we examine how each metric changes across the hyperparameter grid.
We quantify hyperparameter robustness with the coefficient of variation (CV):
\[CV = \frac{\text{std}(a)}{\text{mean}(a)}\]
where \(a\) is the set of values for a given task-agnostic metric across all evaluated hyperparameter settings within a trial. CV is a measure of relative dispersion, which makes it useful for comparing variability across models with different performance scales.

In reservoir computing, hyperparameter sensitivity is commonly analyzed through area-under-curve, performance curves, or robustness intervals rather than through a single standard statistic \parencite{KAWAI201915,jaurigue_timescale_2023,freddi_robust_spiking_2026,srinivasan_ei_balance_2025}. Our use of CV compresses sensitivity into one value per metric: low CV indicates that a metric remains stable across target spectral radius, leak rate, input scaling, and neuron bias, whereas high CV indicates stronger dependence on hyperparameter choice.

\begin{table*}[!t]
    \centering
    \scriptsize
    \caption{Compact summary of the trial definition, hyperparameter sweep, and evaluation pipeline.}
    \label{tab:exp_setup}
    \renewcommand{\arraystretch}{1.15}
    \setlength{\tabcolsep}{4pt}
    \begin{tabular}{p{0.18\textwidth}p{0.53\textwidth}p{0.24\textwidth}}
        \toprule
        Component & Setting & Notes \\
        \midrule

        Trial unit
        & One reservoir realization evaluated across the complete 96-point hyperparameter grid
        & Random quantities defining the realization are held fixed across the grid and redrawn for each trial \\
        
        \addlinespace
        Input projection
        & A scalar input is projected to all \(N=297\) neurons through \(W^{\mathrm{in}}\)
        & The same \(W^{\mathrm{in}}\) realization is used for all 96 hyperparameter combinations within a trial \\
        
        \addlinespace
        MC and IPC sequence
        & 2500 time steps: 500 washout, 1500 training, and 500 testing
        & The same input sequence is used across the hyperparameter grid within a trial \\
        
        \addlinespace
        KR protocol
        & \(M=300\) independent input streams of length \(L=10\); the reservoir is reset before each stream, and the final states are used to compute effective rank
        & KR is evaluated independently of the MC/IPC train--test sequence \\
        
        \addlinespace
        GR protocol
        & \(M=300\) input streams of length \(L=10\), with 7 independently sampled inputs followed by a common 3-step tail; the reservoir is reset before each stream
        & GR is evaluated independently of the MC/IPC train--test sequence \\
        
        \addlinespace
        Readout and capacity metrics
        & MC maximum delay \(D_{\max}=30\); truncated IPC maximum delay \(D_{\max}=30\); IPC Legendre orders \(1\)--\(5\); ridge regularization \(\lambda=10^{-4}\)
        & Pairwise and higher-order cross-delay products are omitted from truncated IPC \\
        
        \addlinespace
        Hyperparameter sweep
        & Target spectral radius \(\{0.6,0.8,0.95,1.05\}\); leak rate \(\{0.6,0.8,1.0\}\); input scaling \(\{0.1,0.5,1.0,1.5\}\); bias magnitude \(\{0.0,0.1\}\)
        & \(4\times3\times4\times2=96\) combinations per trial \\
        
        \addlinespace
        Evaluation pipeline
        & Instantiate reservoir \(\rightarrow\) assign weights/signs \(\rightarrow\) construct fixed trial realization \(\rightarrow\) rescale recurrent matrix to each target spectral radius \(\rightarrow\) evaluate KR, GR, truncated IPC, and MC across the hyperparameter grid \(\rightarrow\) aggregate mean and CV
        & CV is computed across the 96 hyperparameter combinations within each trial before averaging across trials \\
        
        \addlinespace
        Replicates
        & 200 trials per architecture condition
        & Each trial uses a newly drawn reservoir/input realization \\
        
        \bottomrule
    \end{tabular}
\end{table*}

\subsection{Structural perturbations and architecture controls}
To determine which topological and weight-related features influence robustness, we construct controlled network perturbations derived from the \emph{C.~elegans} connectome. Table~\ref{tab:architecture_groups} summarizes the architecture families used in the perturbations. Figure~\ref{fig:shuffle_four_panel} shows representative network perturbations and architecture controls used in this work. Full architecture definitions and visualizations are provided in Supplemental Section~\ref{supp:architectures} (see Supplemental Figures~\ref{fig:a_to_h} and~\ref{fig:i_to_o} for the other architectures).
The raw spectral radius is not held fixed by these constructions; its variation is an outcome of the structural manipulations and is analyzed explicitly below.

\subsection{Experimental setup}
\label{sec:trial_def}
Table~\ref{tab:exp_setup} condenses the trial definition, hyperparameter sweep, and evaluation pipeline.

\section{Results}
\label{sec:results}
The numerical experiments are organized as a sequence of architecture perturbations around the connectome-derived reservoir. The results use the empirical \textit{C.\ elegans} reservoir, as defined in section~\ref{sec:connectome_dataset}. Each architecture is evaluated over the same 96-point hyperparameter grid described in Section~\ref{sec:hyperparameters}. For each trial and task-agnostic metric, mean metric value and CV are calculated across these 96 combinations. We first compare this reservoir against random-topology, binary-weight, and sign-preserving controls. We then examine how reservoir behavior changes across a sweep of E/I edge balances in the empirical connectome. Next, we use degree-preserving connection, sign, and weight shuffles (Table~\ref{tab:architecture_groups}) to compare conditions differing in topology, local sign placement, and weight placement. Finally, we summarize how changes in raw spectral radius and the corresponding normalization factor are associated with changes in performance and CV across these experiments.

\subsection{The empirical connectome occupies a low-variance reservoir regime}
\begin{figure}
    \centering
     \includegraphics[width=\linewidth]{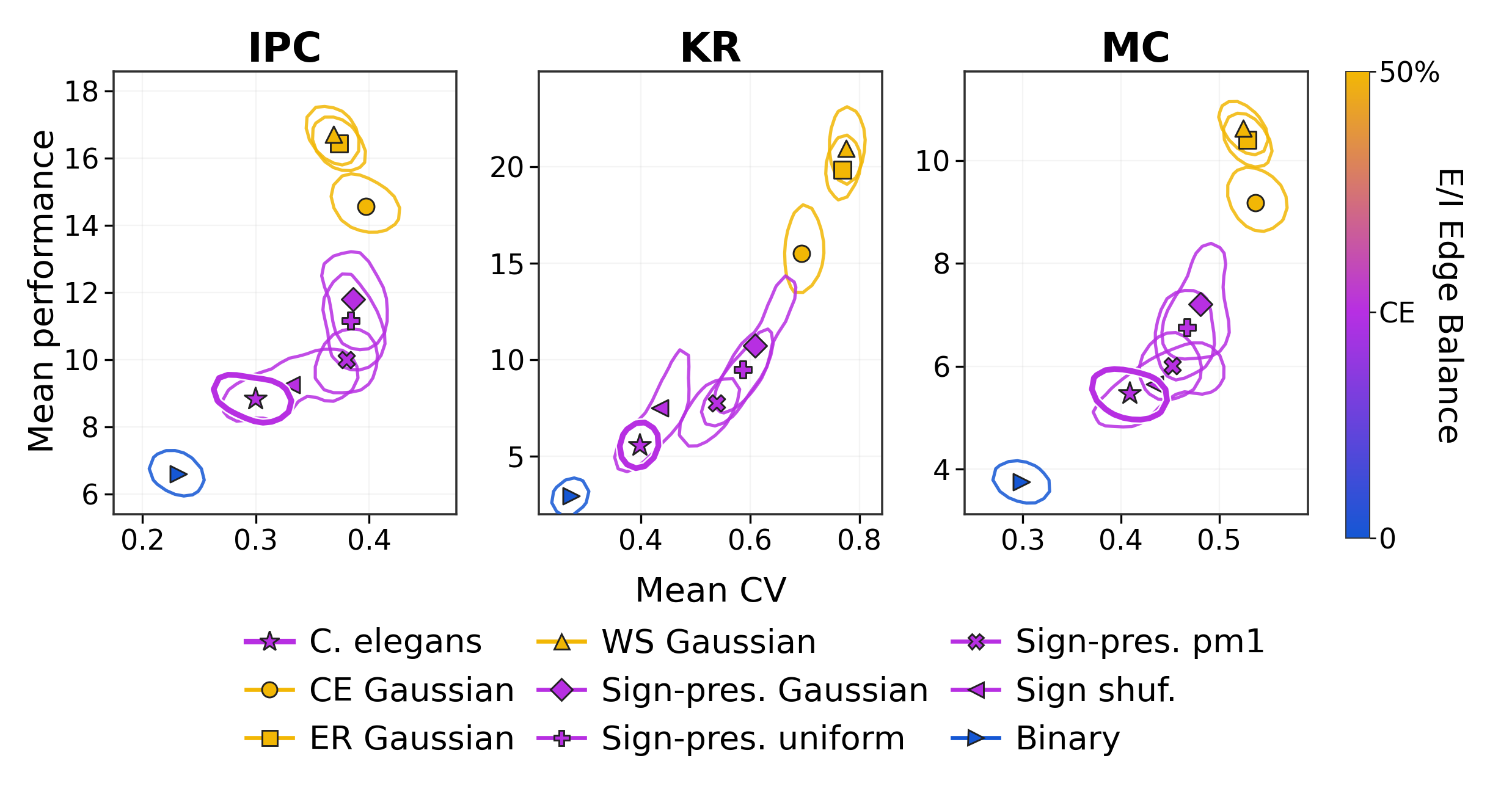}
   \caption{Performance--robustness island plots for the architecture comparison. Each panel shows the mean value of one task-agnostic metric against its hyperparameter CV. Marker position gives the condition mean across trials, and the surrounding contour encloses the highest-density 50\% of trial outcomes. Marker shape identifies the architecture: a star denotes the empirical \textit{C.\ elegans} reservoir; a circle, the \textit{C.\ elegans} topology with Gaussian weights; a square, the edge-matched Erd\H{o}s--R\'enyi network with Gaussian weights; an upward triangle, the Watts--Strogatz network with Gaussian weights; a diamond, sign-preserving Gaussian magnitudes; a plus, sign-preserving uniform magnitudes; a cross, sign-preserving signed (\(\pm1\)) weights; a left-pointing triangle, empirical weight magnitudes with shuffled sign placement; and a right-pointing triangle, all-positive binary weights. Shape and color encode different properties: shape identifies the architecture, whereas color encodes the E/I edge balance. Blue denotes 0\% negative edges, purple denotes the empirical \textit{C.\ elegans} fraction (the ``CE'' tick, approximately 24.25\%), and yellow denotes 50\%. In the legend, CE denotes \textit{C.\ elegans}, ER denotes Erd\H{o}s--R\'enyi, WS denotes Watts--Strogatz, ``Sign-pres.'' denotes a sign-matched control, ``pm1'' denotes weights in \(\{-1,+1\}\), and ``shuf.'' denotes shuffled sign placement. Architecture families and their corresponding manipulations and preserved properties are summarized in Table~\ref{tab:architecture_groups}, with full architecture definitions provided in Appendix~\ref{supp:architectures}.}
    \label{fig:arch}
\end{figure}

Figure~\ref{fig:arch} compares architecture families by mean performance and hyperparameter CV. Each panel shows one metric: truncated IPC, KR, or MC. The vertical axis gives the mean metric value across trials, and the horizontal axis gives the coefficient of variation across the spectral-radius, leak-rate, input-scaling, and neuron-bias sweep. Contours show the 50\% density region for each architecture family, and markers show the corresponding condition means. Color indicates the percentage of edges with negative weights. The 50\% E/I edge balance models, shown in yellow, appear at higher mean performance and higher CV. Models at the empirical connectome's E/I edge balance, shown in purple, appear at lower mean performance and lower CV. The all-positive binary-weight model, shown in blue at 0\% negative edges, appears in the low-performance, low-CV region. Differences in E/I edge balance mark the largest architecture-level separation, while topology, sign placement, and weight magnitudes are associated with additional shifts.

\subsection{E/I edge balance sweeps trace a performance--robustness tradeoff}
\label{balance}
To examine how reservoir behavior varies across different E/I edge balances, we randomly assign synapse signs while leaving the connectivity pattern and weight magnitudes fixed.\footnote{Because signs are assigned randomly, this sweep changes both the E/I edge balance and local sign placement. The sign shuffling control (Sign shuf. in Figure~\ref{fig:arch}, and Supplementary Material~\ref{supp:sign-pres_real_w}) elucidates the effects of this sign placement shuffling.} For each E/I edge balance fraction, we computed the mean value and hyperparameter CV of all four metrics. IPC, KR, and MC are shown here, while GR is analyzed separately in Section~\ref{sec:gr_results}.

\begin{figure} 
    \centering
    \includegraphics[width=1\linewidth]{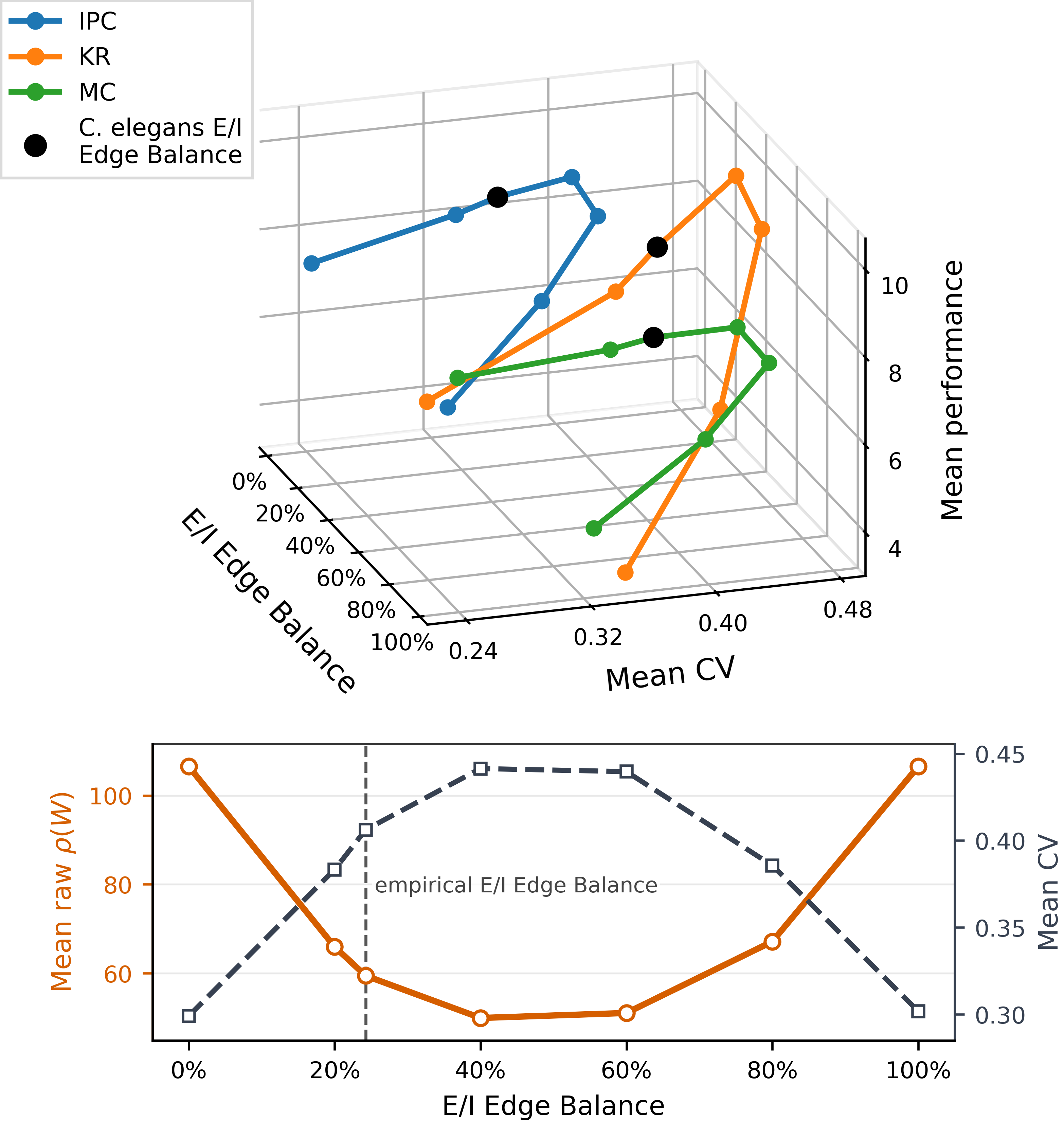}
    \caption{E/I edge balance sweep for the empirical \textit{C.\ elegans} topology. Panel A shows mean task-agnostic performance (vertical axis) and CV (left-right horizontal axis) as functions of the E/I edge balance (front-back horizontal axis). The three black markers in Panel A identify the values of the three task-agnostic metrics at the empirical \textit{C.\ elegans} E/I edge balance. Panel B shows the mean raw spectral radius, \(\rho_{\mathrm{raw}}\), before normalization (solid orange curve with circular markers; left axis) and the mean CV averaged across the three task-agnostic metrics (black dashed curve with square markers; right axis) across the same E/I sweep. The vertical gray dashed line marks the empirical \textit{C.\ elegans} E/I edge balance of approximately \(24.25\%\). Points show averages across 200 trials for each fraction.}
    \label{fig:ambig_cel_frac_curves}
\end{figure}

Figure~\ref{fig:ambig_cel_frac_curves} shows the E/I edge balance sweep for the empirical \textit{C.\ elegans} reservoir. In Panel A, the horizontal axes give the E/I edge balance fraction and mean CV, while the vertical axis gives mean performance. At low and high E/I edge balance fractions, the points lie in a low-performance, low-CV regime. As the E/I edge fraction nears 50\%--toward the middle of the sweep--the mean metric values and CV values both increase. Panel B shows that the raw spectral radius is lowest near the balanced regime and higher toward the all-positive and all-negative endpoints; consequently, the normalization factor, and thus the post-normalization weight scale is larger near the balanced regime. The performance--robustness and raw-spectral-radius trends are replicated when the experiment is rerun with the matched ER topology and the predicted-polarity-only \textit{C.\ elegans} graph in Supplemental Figures~\ref{fig:er_frac_curves} and~\ref{fig:removed_frac}.

\subsection{E/I edge balance, raw spectral radius, and normalization factor vary together}
The E/I edge balance sweep changes not only the balance and placement of positive and negative weights but also the raw spectral radius of the recurrent matrix. As the fraction of negative edges approaches 50\%, \(\rho_{\mathrm{raw}}\) decreases (Figure~\ref{fig:ambig_cel_frac_curves}B; Supplemental Figures~\ref{fig:er_frac_curves}B and~\ref{fig:removed_frac}). Under spectral-radius normalization, \(W^{\mathrm{res}}=\rho_{\mathrm{target}}W'/\rho_{\mathrm{raw}}\), this decrease produces a larger scaling factor and therefore larger weight magnitudes. Sign configuration and post-normalization recurrent-weight scale consequently vary together even though every matrix reaches the same target spectral radius.

The accompanying increases in performance and CV may therefore reflect both changes in sign structure, including local sign placement, and differences in the rescaling required to reach the common target. These contributions are not estimated separately by the present design. The E/I edge balance sweep curves should therefore be interpreted as the result of changing sign ratio under spectral-radius normalization, not as an independently identified effect of sign structure.

\subsection{Topology- and weight-placement comparisons in the performance--robustness plane}
\label{sec:shuf}
\begin{figure}
    \centering
    \includegraphics[width=\linewidth]{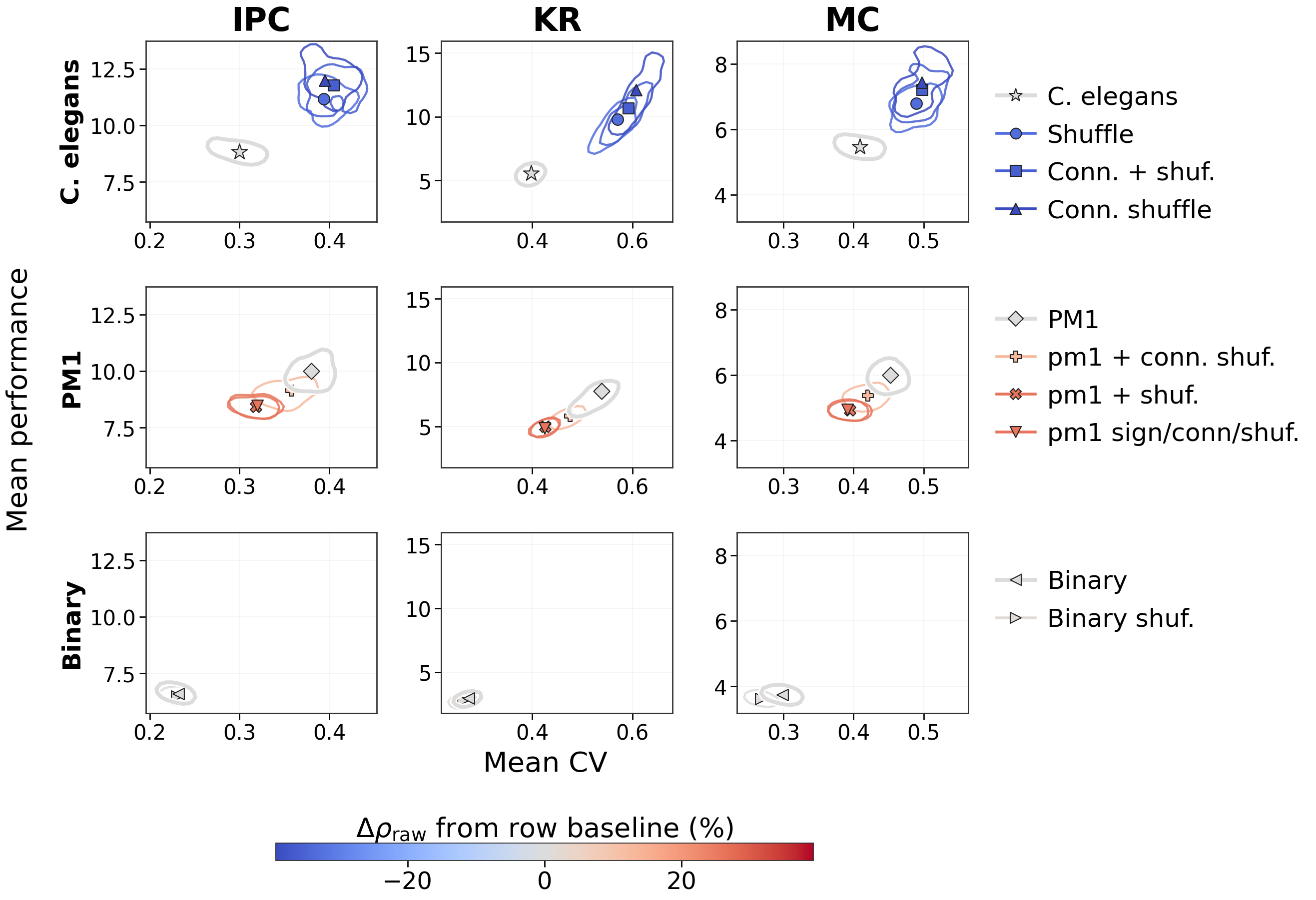}
    \caption{
    Performance--robustness islands for the shuffle controls, shown with 50\% density contours. Columns correspond to truncated IPC, KR, and MC. Rows show real-valued \textit{C.\ elegans} controls, sign-preserving signed-unit controls, and all-positive binary controls. Contour color shows the percent change in raw spectral radius relative to the row baseline, \(\Delta\rho_{\mathrm{raw}}\); blue indicates lower raw spectral radius and red indicates higher raw spectral radius. Exact legend-label mappings, abbreviations, and full definitions for all architectures shown here are provided in Supplementary Material~\ref{supp:architectures}.}
    \label{fig:shuf_all}
\end{figure}

Figure~\ref{fig:shuf_all} places these controls in the performance-robustness plane used above. Columns correspond to the three task-agnostic metrics defined in Sec.~\ref{sec:methods}. Rows separate three baseline weight regimes: real-valued \textit{C.\ elegans} weights on the top row, signed-unit weights in the middle row, and all-positive binary weights on the bottom row. Contour color gives the percent change in raw spectral radius relative to the row baseline, with blue indicating lower \(\rho_{\mathrm{raw}}\) and red indicating higher \(\rho_{\mathrm{raw}}\).

In the first row, the shuffle controls move away from the unshuffled predicted-polarity reservoir toward higher mean performance and higher CV. These contours and symbols are blue, indicating lower raw spectral radius relative to the row baseline (gray star). In the signed-unit row, the sign- and topology-shuffled controls lie at lower performance and lower CV than the signed-unit baseline, and their contours and symbols are red, indicating a higher raw spectral radius relative to the baseline. In the all-positive binary row, the shuffled controls remain closer to the row baseline and show smaller shifts in the performance--CV plane. The paired CV differences for these rows are reported in Supplemental Tables~\ref{tab:shuffle_cv_real}--\ref{tab:shuffle_cv_binary}. Across weight regimes, these controls show that changes in local topology and weight placement are accompanied by shifts in both performance and robustness, with these shifts inversely related to changes in raw spectral radius.

\subsection{Raw spectral radius tracks the performance--robustness tradeoff}
\label{sec:raw_rho_summary}
\begin{figure}
    \centering
    \includegraphics[width=\linewidth]{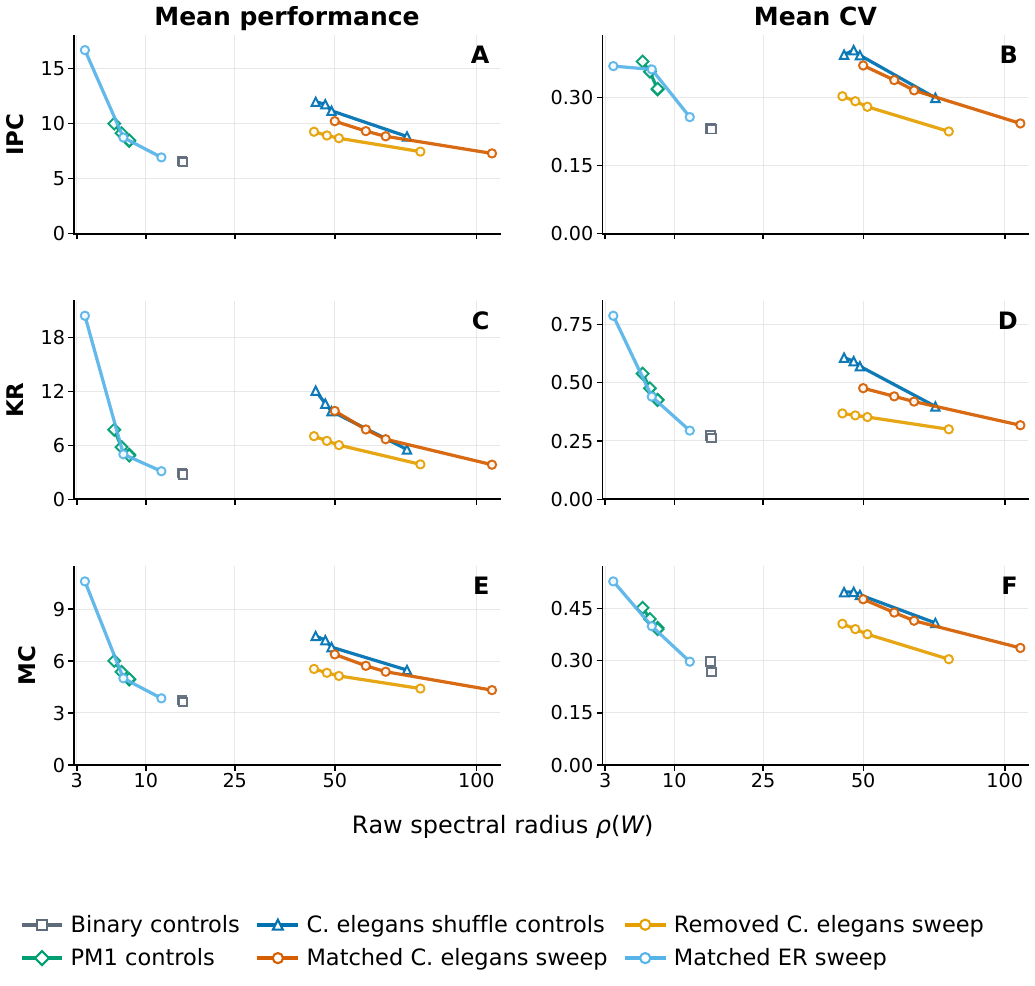}
    \caption{Relationship between raw spectral radius and reservoir behavior across E/I edge balance sweep and shuffle-control families under spectral-radius normalization. Rows correspond to truncated IPC, KR, and MC. The left column shows mean task-agnostic performance as a function of raw spectral radius before normalization, \(\rho_{\mathrm{raw}}\), on a log scale; the right column shows the corresponding hyperparameter CV. Lines connect points within each experimental family. The binary, PM1, and \textit{C.\ elegans} shuffle-controls are from Figure~\ref{fig:shuf_all}; the \textit{C.\ elegans} E/I sweep, restricted to conditions with less than \(50\%\) negative edges, is from Figure~\ref{fig:ambig_cel_frac_curves}.}
    \label{fig:raw_rho_summary}
\end{figure}

Structural perturbations change the raw spectral radius and therefore the global scaling factor required to reach a common target radius. Figure~\ref{fig:raw_rho_summary} summarizes the resulting associations across the E/I edge balance sweeps and shuffle controls. Across the three metrics, lower \(\rho_{\mathrm{raw}}\) coincides with higher mean performance and higher CV, whereas higher \(\rho_{\mathrm{raw}}\) coincides with lower performance and lower CV. The same direction appears in multiple tested families: E/I edge balance sweeps, signed-unit controls, binary topology controls, and real-weight shuffles.

Thus, the synthesis is not simply that the E/I edge balance matters.  Controls with the same E/I edge balance occupy different performance--robustness regimes when topology, local sign placement, weight-magnitude distribution, or magnitude placement differs. At the same time, matrix structure and the normalization factor vary together in these comparisons.

\subsection{Generalization rank extends the performance--robustness tradeoff}
\label{sec:gr_results}

\begin{figure*}[t]
    \centering
    \includegraphics[
        width=\textwidth
    ]{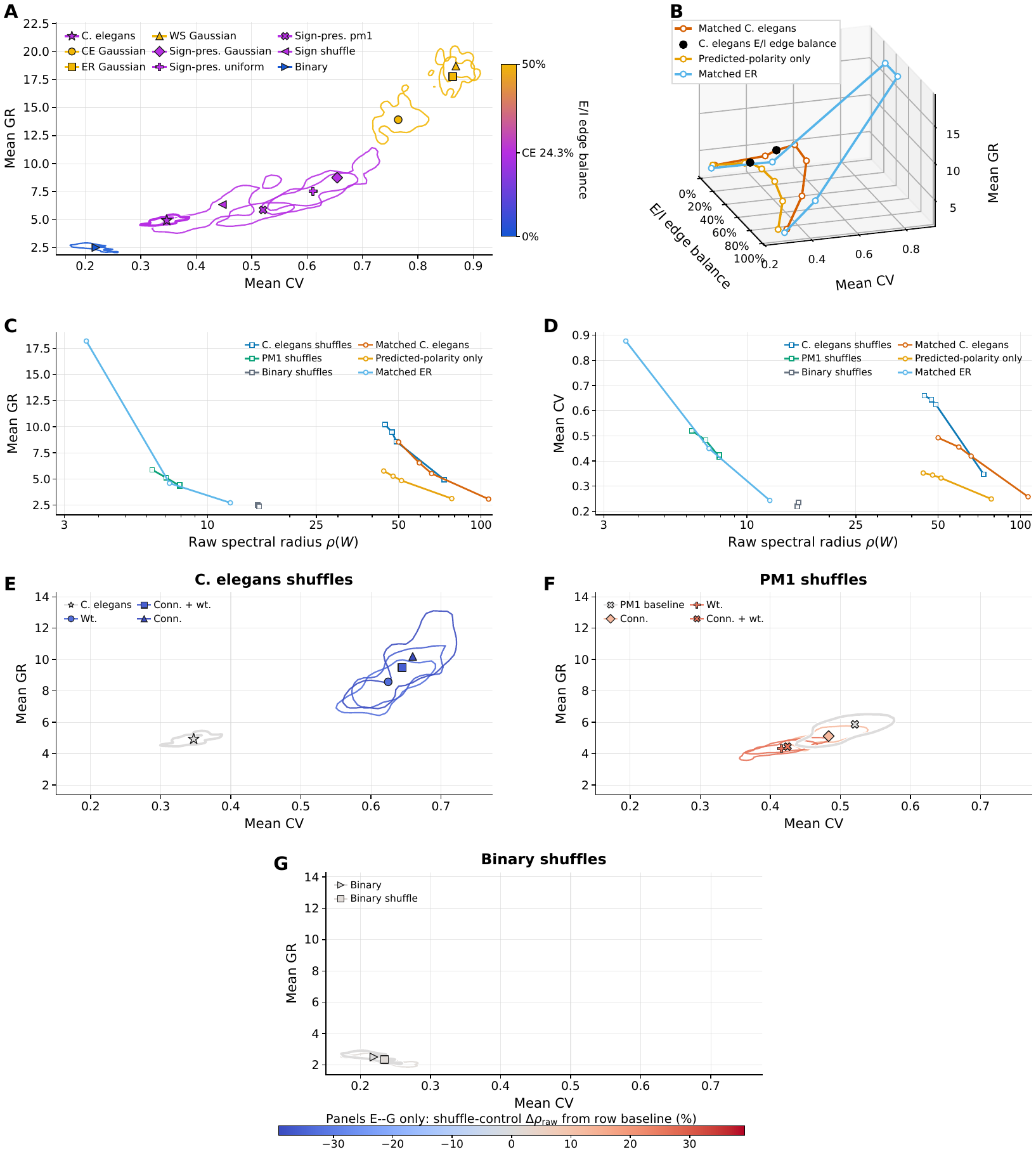}
    \caption{Lower generalization rank (GR) indicates better generalization.
    \textbf{(A)} Mean GR versus Mean CV across architectures. Contours enclose the highest-density 50\% of each distribution, and color denotes E/I edge balance.
    \textbf{(B)} Mean CV and Mean GR across the full E/I edge balance sweeps for the matched \textit{C. elegans}, predicted-polarity-only, and matched-ER networks; black markers indicate the empirical \textit{C. elegans} edge balance.
    \textbf{(C--D)} Mean GR and Mean CV, respectively, as functions of raw spectral radius, with E/I edge balance sweep conditions shown through 50\% E/I edge balance.
    \textbf{(E--G)} Mean GR versus Mean CV for the \textit{C. elegans}, PM1, and binary shuffle controls. Contours summarize the corresponding distributions, while color in panels E--G indicates the percentage change in raw spectral radius relative to each row's baseline.
    Across most conditions, mean GR and the hyperparameter CV of GR increase together, indicating losses of both generalization and hyperparameter robustness.}
    \label{fig:generalization_rank_summary}
\end{figure*}

Generalization rank was analyzed separately because lower GR indicates better common-tail generalization. Figure~\ref{fig:generalization_rank_summary} shows that conditions associated with higher IPC, KR, and MC also tend to have higher GR and greater hyperparameter CV of GR, particularly when raw spectral radius is lower. The shuffle controls show the same general pattern. This architecture ordering was largely preserved across the tested common-tail lengths \(L_c\in\{1,2,3,4,5,7,10\}\) (Supplemental Section~\ref{supp:gr_tail_sensitivity}).

Thus, increases in performance are accompanied by losses in two forms of robustness: greater sensitivity to hyperparameter choice and poorer common-tail generalization. 

\section{Discussion}
This study uses an empirical \textit{C.\ elegans} connectome \parencite{fenyves_szilágyi_zsolt_vassy_csaba_2020} reservoir as a recurrent neural network model for asking how architecture relates to hyperparameter sensitivity. The empirical \textit{C.\ elegans} connectome's performance varies relatively little across the hyperparameter grid. Among the tested perturbations, the E/I edge balance sweeps produce the largest shifts in task-agnostic performance and CV under spectral-radius normalization. The magnitude and direction of these shifts nevertheless depend on topology, sign placement, the distribution and placement of weight magnitudes, and the normalization factor determined by the resulting raw spectral radius. E/I edge balance should therefore be interpreted as an architecture-dependent parameter rather than an independent determinant of reservoir dynamics.

Figure~\ref{fig:arch} places the empirical \textit{C.\ elegans} model in a relatively low-CV regime across the three task-agnostic performance metrics. Furthermore, weight shuffling, connection shuffling, and connection-plus-weight shuffling all move the reservoir toward higher mean performance and higher CV (Figure~\ref{fig:shuf_all}, top row). Thus, in this reservoir abstraction, the local topology, sign placement, and weight placement of the empirical \textit{C.\ elegans} model are associated with lower hyperparameter sensitivity and better generalization across input histories, rather than maximal computational performance.

The E/I edge balance sweeps show the core performance--robustness tradeoff. Figure~\ref{fig:ambig_cel_frac_curves} in the main text and Supplemental Figures~\ref{fig:er_frac_curves} and~\ref{fig:removed_frac} show that configurations nearer 50\% negative edges generally have higher mean task-agnostic performance and higher CV. The same qualitative trend is observed across all architectures tested. Under the present modeling and normalization choices, the empirical \textit{C.\ elegans} instantiation therefore lies on the low-variance side of this curve rather than near the peak-performance region. The sign imbalance observed in the labeled connections is associated with a more hyperparameter-invariant, lower-performance operating point.

The GR analysis broadens the performance--robustness tradeoff identified by the other metrics. The conditions with higher IPC, KR, and MC also tend to have both higher GR and higher mean CV. Because lower GR indicates better common-tail generalization, increased performance is therefore accompanied by losses in two forms of robustness: greater sensitivity to hyperparameter choice and poorer common-tail generalization.

The positive association between MC and GR is compatible with a memory--generalization tradeoff: MC increases when reservoir states retain information about earlier inputs, enabling their reconstruction, whereas common-tail GR treats persistent dependence on that history as reduced generalization.

In the E/I edge balance sweeps, more balanced configurations have lower \(\rho_{\mathrm{raw}}\) and therefore receive greater global amplification when rescaled to a common target radius. The associated changes in performance, CV, reservoir activity, and normalized recurrent-weight magnitudes consequently describe the joint result of changing sign structure and changing the weight scale required by normalization. The sign-balanced reservoir is active more often, and its neurons become active more often, than those in the all-positive reservoir (Supplemental Figure~\ref{fig:activity_0_vs_50}). The mean absolute normalized weight within closed three-neuron subgraphs also increases toward the middle of the E/I edge balance sweep (Supplemental Figure~\ref{fig:triad_sign_fraction}). Greater recurrent-weight amplification may enrich separable reservoir states while also preserving differences in earlier input history, providing a possible explanation for higher performance and poorer generalization.

These analyses are consistent with the algebraic coupling between raw radius and the normalization factor, but they do not separate its components. Accordingly, the present design cannot determine how much of the observed change is attributable to sign, weight or topological arrangement independently of its consequences on raw spectral radius.

This normalization-based account also connects the present results to recent reservoir work on the no-strong-loops principle. Hadaeghi and colleagues show that stronger reciprocal loops inflate \(\rho(W)\), forcing spectral-radius normalization to downscale recurrent weights more strongly and reducing memory capacity and kernel rank \parencite{hadaeghi_no_strong_loops_2025}.

The E/I edge balance sweeps reproduce much of the main architecture-level separation, but they do not account for differences among controls that share the same global sign ratio. The degree-preserving shuffle experiments reveal additional differences: controls with lower \(\rho_{\mathrm{raw}}\) tend to move toward higher mean performance and higher CV, whereas controls with higher \(\rho_{\mathrm{raw}}\) tend to move in the opposite direction (Figure~\ref{fig:shuf_all}).

Direct biological interpretation should be cautious because the connectome model compresses substantial uncertainty into a static network with predicted chemical polarity labels. Edges use EleganSign-predicted polarity, but polarity is inferred rather than directly measured: they come from gene-expression and neurotransmitter logic, not from voltage- or calcium-response measurements \parencite{fenyves_szilágyi_zsolt_vassy_csaba_2020}. The model includes chemical synapses only and omits electrical synapses, making it a less complete representation of the biological circuit. Coverage is incomplete as well: many chemical synapses have complex or unpredicted labels, task--agnostic metrics depend on how those cases are handled. Furthermore, \textit{C.\ elegans} E/I edge balance varies across development and aging \parencite{wirak_florman_alkema_connor_gabel_2022,tanis_bellemer_moresco_forbush_koelle_2009}, and our model delivers inputs to all neurons rather than to biologically separated populations, such as sending inputs to sensory neurons and taking outputs from motor neurons or muscles. Finally, changes in cellular E/I composition can be accompanied by compensatory changes in connection number \parencite{sukenik_connection_numbers_2021}, suggesting that E/I composition and network structure are biologically coupled. Thus, this work should be read as a structural reservoir analysis, not a faithful dynamical simulation of the animal.

The shape of the sign-balance curves in Figure~\ref{fig:ambig_cel_frac_curves} should also be interpreted through the reservoir abstraction. The state nonlinearity is \(\tanh\), which is odd-symmetric, and the readout is linear. Under these choices, exchanging positive recurrent weights for negative recurrent weights changes the sign structure of the recurrent drive, but the linear readout can absorb some corresponding sign changes in the state features. These modeling choices can therefore favor approximately symmetric E/I edge balance sweep curves, although the recurrent dynamics, leak, bias, and topology do not guarantee exact symmetry.

The odd-symmetric \(\tanh\) nonlinearity and linear readout used here help explain why the present curves differ from studies in which the best operating point is slightly inhibition-dominated. \cite{srinivasan_ei_balance_2025} as well as \cite{de_graaf_spinal_inhibition_2025} report performance benefits from balanced or slightly inhibition-dominated adaptive E/I regimes. We hypothesize that the difference may partly reflect the activation functions used across these models: Srinivasan et al use a sigmoid activation, whereas de Graaf et al use a rectified \(\tanh\) activation. In such models, the activation function is not fully sign-symmetric. Positive and negative inputs are transformed differently, so the optimal E/I ratio can shift away from exactly 50:50.

For reservoir computing and recurrent neural network design, raw spectral radius should therefore be reported alongside architecture comparisons: it determines the global scaling required by spectral-radius normalization and changes alongside performance, generalization, and hyperparameter sensitivity, but it does not fully explain any of them. 

\section{Future directions}
First, the same framework could be used to investigate how recurrent network optimization algorithms alter task-agnostic statistics, and whether task-driven optimization moves structured reservoirs along the same performance--robustness axes identified here. Relevant comparisons include reward-modulated reinforcement learning in densely recurrent biological networks \parencite{churchland_garcia-ojalvo_2026}, covariance-matrix-adaptation evolution strategies (CMA-ES) \parencite{hansen_cmaes_2016}, and scalable, derivative-free evolution strategies such as OpenAI-ES \parencite{salimans_evolution_2017}.

Second, driving and reading out from all neurons provides a controlled, task-agnostic characterization of the complete reservoir without imposing a particular biological input--output partition. A complementary extension could restrict inputs to sensory neurons and define outputs over motor neurons or muscles, bringing the task closer to biological information flow through the \textit{C.\ elegans} circuit and enabling more direct comparison with previous connectome-based computational models \parencite{KAWAI201915,yu_bpu_2025}.

Third, the uncertainty in synaptic-sign assignment should be treated as an experimental variable in its own right. Comparing alternative neurotransmitter-to-sign mappings, and systematically varying the treatment of ambiguous edges, would reveal whether the observed E/I-balance patterns persist across plausible E/I edge balances.

Fourth, future work should investigate which hyperparameters contribute most strongly to variability in each task-agnostic metric. Different metrics may be sensitive to different aspects of the reservoir dynamics; for example, memory capacity may vary little with one hyperparameter but respond strongly to another. Examining hyperparameter-specific variance would reveal which parameters each metric is most sensitive to and whether those sensitivities differ across metrics.

\section*{Acknowledgments}
The authors thank Anne Churchland, Michael Brodesky, Remi Cannon, and Autumn Aabram, for their support and for conversations that helped clarify the ideas developed in this work.

\section*{Code and Data Availability}

Code, analysis materials, and manuscript sources are available at \url{https://github.com/mchurchland/cel_res_hypr}.

\section*{AI Usage Disclosure}

Artificial intelligence tools, specifically ChatGPT (OpenAI, GPT-5 series) and Codex (OpenAI), were used as assistive tools during the course of this work.

AI tools were used to review text for clarity, grammar, and structure, and to identify ambiguous or overstated claims. All text included in this manuscript was written, reviewed, and edited by the authors, and any AI-assisted edits were manually verified and revised as necessary to ensure accuracy.

AI tools were also used to assist in identifying relevant literature. These tools were used in a search-assistance capacity to locate papers related to reservoir computing and recurrent biological networks. All cited papers were independently located, read, and verified by the author prior to inclusion.

In addition, AI tools were used as a programming reference to identify relevant libraries, functions, and debugging strategies. They were also used to assist in generating example plotting code and suggesting implementation approaches. All code used in this project was written, reviewed, and validated by the author.

AI tools were used solely as assistive instruments. All scientific claims, analysis, interpretations, and conclusions presented in this work are the independent work of the authors.

\clearpage
\bibliographystyle{style_neural_networks/cas-model2-names}
\bibliography{bibliography}

\begin{thebibliography}{41}
\expandafter\ifx\csname natexlab\endcsname\relax\def\natexlab#1{#1}\fi
\providecommand{\url}[1]{\texttt{#1}}
\providecommand{\href}[2]{#2}
\providecommand{\path}[1]{#1}
\providecommand{\DOIprefix}{doi:}
\providecommand{\ArXivprefix}{arXiv:}
\providecommand{\URLprefix}{URL: }
\providecommand{\Pubmedprefix}{pmid:}
\providecommand{\doi}[1]{\href{http://dx.doi.org/#1}{\path{#1}}}
\providecommand{\Pubmed}[1]{\href{pmid:#1}{\path{#1}}}
\providecommand{\bibinfo}[2]{#2}
\ifx\xfnm\relax \def\xfnm[#1]{\unskip,\space#1}\fi
\bibitem[{Barab{\'a}si(2016)}]{network_science}
\bibinfo{author}{Barab{\'a}si, A.L.}, \bibinfo{year}{2016}.
\newblock \bibinfo{title}{Network Science}.
\newblock \bibinfo{publisher}{Cambridge University Press}, \bibinfo{address}{Cambridge}.
\newblock \URLprefix \url{https://www.cambridge.org/core/books/network-science/E7FD32714480E1E7AA1BCED4E720AF96}.
\bibitem[{B{\"u}sing et~al.(2010)B{\"u}sing, Schrauwen and Legenstein}]{busing_connectivity_2010}
\bibinfo{author}{B{\"u}sing, L.}, \bibinfo{author}{Schrauwen, B.}, \bibinfo{author}{Legenstein, R.}, \bibinfo{year}{2010}.
\newblock \bibinfo{title}{Connectivity, dynamics, and memory in reservoir computing with binary and analog neurons}.
\newblock \bibinfo{journal}{Neural Computation} \bibinfo{volume}{22}, \bibinfo{pages}{1272--1311}.
\newblock \DOIprefix\doi{10.1162/neco.2009.01-09-947}.
\bibitem[{Casal et~al.(2020)Casal, Galella, Vilarroya and Garcia-Ojalvo}]{Casal2020}
\bibinfo{author}{Casal, M.A.}, \bibinfo{author}{Galella, S.}, \bibinfo{author}{Vilarroya, O.}, \bibinfo{author}{Garcia-Ojalvo, J.}, \bibinfo{year}{2020}.
\newblock \bibinfo{title}{Soft-wired long-term memory in a natural recurrent neuronal network}.
\newblock \bibinfo{journal}{Chaos: An Interdisciplinary Journal of Nonlinear Science} \bibinfo{volume}{30}, \bibinfo{pages}{061101}.
\newblock \DOIprefix\doi{10.1063/5.0009709}.
\bibitem[{Churchland and Garcia-Ojalvo(2026)}]{churchland_garcia-ojalvo_2026}
\bibinfo{author}{Churchland, M.W.}, \bibinfo{author}{Garcia-Ojalvo, J.}, \bibinfo{year}{2026}.
\newblock \bibinfo{title}{Reinforcement learning in densely recurrent biological networks}.
\newblock \bibinfo{journal}{iScience} \bibinfo{volume}{29}, \bibinfo{pages}{114436}.
\newblock \DOIprefix\doi{10.1016/j.isci.2025.114436}.
\bibitem[{Cook et~al.(2019)Cook, Jarrell, Brittin, Wang, Bloniarz, Yakovlev, Nguyen, Tang, Bayer, Duerr, B{\"u}low, Hobert, Hall and Emmons}]{Cook2019}
\bibinfo{author}{Cook, S.J.}, \bibinfo{author}{Jarrell, T.A.}, \bibinfo{author}{Brittin, C.A.}, \bibinfo{author}{Wang, Y.}, \bibinfo{author}{Bloniarz, A.E.}, \bibinfo{author}{Yakovlev, M.A.}, \bibinfo{author}{Nguyen, K.C.Q.}, \bibinfo{author}{Tang, L.T.H.}, \bibinfo{author}{Bayer, E.A.}, \bibinfo{author}{Duerr, J.S.}, \bibinfo{author}{B{\"u}low, H.E.}, \bibinfo{author}{Hobert, O.}, \bibinfo{author}{Hall, D.H.}, \bibinfo{author}{Emmons, S.W.}, \bibinfo{year}{2019}.
\newblock \bibinfo{title}{Whole-animal connectomes of both \emph{Caenorhabditis elegans} sexes}.
\newblock \bibinfo{journal}{Nature} \bibinfo{volume}{571}, \bibinfo{pages}{63--71}.
\newblock \DOIprefix\doi{10.1038/s41586-019-1352-7}.
\bibitem[{Dambre et~al.(2012)Dambre, Verstraeten, Schrauwen and Massar}]{dambre_ipc_2012}
\bibinfo{author}{Dambre, J.}, \bibinfo{author}{Verstraeten, D.}, \bibinfo{author}{Schrauwen, B.}, \bibinfo{author}{Massar, S.}, \bibinfo{year}{2012}.
\newblock \bibinfo{title}{Information processing capacity of dynamical systems}.
\newblock \bibinfo{journal}{Scientific Reports} \bibinfo{volume}{2}, \bibinfo{pages}{514}.
\newblock \DOIprefix\doi{10.1038/srep00514}.
\bibitem[{{EleganSign}(2026)}]{elegansign_web}
\bibinfo{author}{{EleganSign}}, \bibinfo{year}{2026}.
\newblock \bibinfo{title}{Elegansign}.
\newblock \URLprefix \url{https://elegansign.linkgroup.hu/}. \bibinfo{note}{web application with no publication date; accessed June 26, 2026}.
\bibitem[{Fenyves et~al.(2020)Fenyves, Szil{\'a}gyi, Vassy, S{\H{o}}ti and Csermely}]{fenyves_szilágyi_zsolt_vassy_csaba_2020}
\bibinfo{author}{Fenyves, B.G.}, \bibinfo{author}{Szil{\'a}gyi, G.S.}, \bibinfo{author}{Vassy, Z.}, \bibinfo{author}{S{\H{o}}ti, C.}, \bibinfo{author}{Csermely, P.}, \bibinfo{year}{2020}.
\newblock \bibinfo{title}{Synaptic polarity and sign-balance prediction using gene expression data in the \emph{Caenorhabditis elegans} chemical synapse neuronal connectome network}.
\newblock \bibinfo{journal}{PLOS Computational Biology} \bibinfo{volume}{16}, \bibinfo{pages}{e1007974}.
\newblock \DOIprefix\doi{10.1371/journal.pcbi.1007974}.
\bibitem[{Freddi et~al.(2026)Freddi, Seseri, Nigrisoli and Basti}]{freddi_robust_spiking_2026}
\bibinfo{author}{Freddi, R.}, \bibinfo{author}{Seseri, N.}, \bibinfo{author}{Nigrisoli, D.}, \bibinfo{author}{Basti, A.}, \bibinfo{year}{2026}.
\newblock \bibinfo{title}{Bridging theory and practice in crafting robust spiking reservoirs}.
\newblock \DOIprefix\doi{10.48550/arXiv.2604.06395}, \href{http://arxiv.org/abs/2604.06395}{\tt arXiv:2604.06395}.
\bibitem[{Gabalda-Sagarra et~al.(2018)Gabalda-Sagarra, Carey and Garcia-Ojalvo}]{gabalda2018}
\bibinfo{author}{Gabalda-Sagarra, M.}, \bibinfo{author}{Carey, L.B.}, \bibinfo{author}{Garcia-Ojalvo, J.}, \bibinfo{year}{2018}.
\newblock \bibinfo{title}{Recurrence-based information processing in gene regulatory networks}.
\newblock \bibinfo{journal}{Chaos: An Interdisciplinary Journal of Nonlinear Science} \bibinfo{volume}{28}, \bibinfo{pages}{106313}.
\newblock \DOIprefix\doi{10.1063/1.5039861}.
\bibitem[{Gallicchio(2019)}]{gallicchio_chasing_2019}
\bibinfo{author}{Gallicchio, C.}, \bibinfo{year}{2019}.
\newblock \bibinfo{title}{Chasing the echo state property}, in: \bibinfo{booktitle}{ESANN 2019 Proceedings, European Symposium on Artificial Neural Networks, Computational Intelligence and Machine Learning}, \bibinfo{publisher}{i6doc.com}, \bibinfo{address}{Bruges, Belgium}. pp. \bibinfo{pages}{667--672}.
\newblock \URLprefix \url{https://www.esann.org/sites/default/files/proceedings/legacy/es2019-76.pdf}.
\bibitem[{de~Graaf et~al.(2025)de~Graaf, Wagner, Mochizuki and Le~Mouel}]{de_graaf_spinal_inhibition_2025}
\bibinfo{author}{de~Graaf, M.L.}, \bibinfo{author}{Wagner, H.}, \bibinfo{author}{Mochizuki, L.}, \bibinfo{author}{Le~Mouel, C.}, \bibinfo{year}{2025}.
\newblock \bibinfo{title}{Decreased spinal inhibition leads to undiversified locomotor patterns}.
\newblock \bibinfo{journal}{Biological Cybernetics} \bibinfo{volume}{119}, \bibinfo{pages}{12}.
\newblock \DOIprefix\doi{10.1007/s00422-025-01011-7}.
\bibitem[{Hadaeghi et~al.(2025)Hadaeghi, Fakhar, Khajehnejad and Hilgetag}]{hadaeghi_no_strong_loops_2025}
\bibinfo{author}{Hadaeghi, F.}, \bibinfo{author}{Fakhar, K.}, \bibinfo{author}{Khajehnejad, M.}, \bibinfo{author}{Hilgetag, C.C.}, \bibinfo{year}{2025}.
\newblock \bibinfo{title}{A computational perspective on the no-strong-loops principle in brain networks}.
\newblock \bibinfo{howpublished}{bioRxiv preprint}.
\newblock \DOIprefix\doi{10.1101/2025.09.24.678310}. \bibinfo{note}{version 1, posted September 25, 2025}.
\bibitem[{Hansen(2016)}]{hansen_cmaes_2016}
\bibinfo{author}{Hansen, N.}, \bibinfo{year}{2016}.
\newblock \bibinfo{title}{The {CMA} evolution strategy: A tutorial}.
\newblock \DOIprefix\doi{10.48550/arXiv.1604.00772}, \href{http://arxiv.org/abs/1604.00772}{\tt arXiv:1604.00772}. \bibinfo{note}{version 2, revised March 10, 2023}.
\bibitem[{Iannello et~al.(2026)Iannello, Ciampi, Lagani, Tonelli, Crocco, Calcagnile, Di~Garbo, Cremisi and Amato}]{iannello_brc_2025}
\bibinfo{author}{Iannello, L.}, \bibinfo{author}{Ciampi, L.}, \bibinfo{author}{Lagani, G.}, \bibinfo{author}{Tonelli, F.}, \bibinfo{author}{Crocco, E.}, \bibinfo{author}{Calcagnile, L.M.}, \bibinfo{author}{Di~Garbo, A.}, \bibinfo{author}{Cremisi, F.}, \bibinfo{author}{Amato, G.}, \bibinfo{year}{2026}.
\newblock \bibinfo{title}{From neurons to computation: Biological reservoir computing for pattern recognition}, in: \bibinfo{editor}{Taniguchi, T.}, \bibinfo{editor}{Leung, C.S.A.}, \bibinfo{editor}{Kozuno, T.}, \bibinfo{editor}{Yoshimoto, J.}, \bibinfo{editor}{Mahmud, M.}, \bibinfo{editor}{Doborjeh, M.}, \bibinfo{editor}{Doya, K.} (Eds.), \bibinfo{booktitle}{Neural Information Processing: 32nd International Conference, {ICONIP} 2025, Proceedings, Part V}, \bibinfo{publisher}{Springer}, \bibinfo{address}{Singapore}. pp. \bibinfo{pages}{114--131}.
\newblock \DOIprefix\doi{10.1007/978-981-95-4100-3_9}. \bibinfo{note}{first published online November 20, 2025}.
\bibitem[{Jaeger(2001)}]{jaeger_echo_state_2001}
\bibinfo{author}{Jaeger, H.}, \bibinfo{year}{2001}.
\newblock \bibinfo{title}{The ``Echo State'' Approach to Analysing and Training Recurrent Neural Networks}.
\newblock \bibinfo{type}{Technical Report} \bibinfo{number}{148}. GMD -- German National Research Center for Information Technology.
\newblock \URLprefix \url{https://www.ai.rug.nl/minds/uploads/EchoStatesTechRep.pdf}. \bibinfo{note}{corrected version published in 2010}.
\bibitem[{Jaeger(2002)}]{Jaeger2002STMESN}
\bibinfo{author}{Jaeger, H.}, \bibinfo{year}{2002}.
\newblock \bibinfo{title}{Short Term Memory in Echo State Networks}.
\newblock \bibinfo{type}{GMD Report} \bibinfo{number}{152}. GMD -- Forschungszentrum Informationstechnik GmbH.
\newblock \URLprefix \url{https://www.ai.rug.nl/minds/uploads/STMEchoStatesTechRep.pdf}.
\bibitem[{Jaurigue and L{\"u}dge(2024)}]{jaurigue_timescale_2023}
\bibinfo{author}{Jaurigue, L.}, \bibinfo{author}{L{\"u}dge, K.}, \bibinfo{year}{2024}.
\newblock \bibinfo{title}{Reducing reservoir computer hyperparameter dependence by external timescale tailoring}.
\newblock \bibinfo{journal}{Neuromorphic Computing and Engineering} \bibinfo{volume}{4}, \bibinfo{pages}{014001}.
\newblock \DOIprefix\doi{10.1088/2634-4386/ad1d32}.
\bibitem[{Kawai et~al.(2019)Kawai, Park and Asada}]{KAWAI201915}
\bibinfo{author}{Kawai, Y.}, \bibinfo{author}{Park, J.}, \bibinfo{author}{Asada, M.}, \bibinfo{year}{2019}.
\newblock \bibinfo{title}{A small-world topology enhances the echo state property and signal propagation in reservoir computing}.
\newblock \bibinfo{journal}{Neural Networks} \bibinfo{volume}{112}, \bibinfo{pages}{15--23}.
\newblock \DOIprefix\doi{10.1016/j.neunet.2019.01.002}.
\bibitem[{Legenstein and Maass(2007)}]{legenstein_maass_2007}
\bibinfo{author}{Legenstein, R.}, \bibinfo{author}{Maass, W.}, \bibinfo{year}{2007}.
\newblock \bibinfo{title}{Edge of chaos and prediction of computational performance for neural circuit models}.
\newblock \bibinfo{journal}{Neural Networks} \bibinfo{volume}{20}, \bibinfo{pages}{323--334}.
\newblock \DOIprefix\doi{10.1016/j.neunet.2007.04.017}.
\bibitem[{Love et~al.(2023)Love, Msiska, Mulkers, Bourianoff, Leliaert and Everschor-Sitte}]{love_task_agnostic_2021}
\bibinfo{author}{Love, J.}, \bibinfo{author}{Msiska, R.}, \bibinfo{author}{Mulkers, J.}, \bibinfo{author}{Bourianoff, G.}, \bibinfo{author}{Leliaert, J.}, \bibinfo{author}{Everschor-Sitte, K.}, \bibinfo{year}{2023}.
\newblock \bibinfo{title}{Spatial analysis of physical reservoir computers}.
\newblock \bibinfo{journal}{Physical Review Applied} \bibinfo{volume}{20}, \bibinfo{pages}{044057}.
\newblock \DOIprefix\doi{10.1103/PhysRevApplied.20.044057}.
\bibitem[{Luko{\v{s}}evi{\v{c}}ius(2012)}]{Lukosevicius2012PracticalESN}
\bibinfo{author}{Luko{\v{s}}evi{\v{c}}ius, M.}, \bibinfo{year}{2012}.
\newblock \bibinfo{title}{A practical guide to applying echo state networks}, in: \bibinfo{editor}{Montavon, G.}, \bibinfo{editor}{Orr, G.B.}, \bibinfo{editor}{M{\"u}ller, K.R.} (Eds.), \bibinfo{booktitle}{Neural Networks: Tricks of the Trade}. \bibinfo{publisher}{Springer}, \bibinfo{address}{Berlin, Heidelberg}. volume \bibinfo{volume}{7700} of \textit{\bibinfo{series}{Lecture Notes in Computer Science}}, pp. \bibinfo{pages}{659--686}.
\newblock \DOIprefix\doi{10.1007/978-3-642-35289-8_36}.
\bibitem[{Maass et~al.(2004)Maass, Legenstein and Bertschinger}]{maass_comp_power_2004}
\bibinfo{author}{Maass, W.}, \bibinfo{author}{Legenstein, R.}, \bibinfo{author}{Bertschinger, N.}, \bibinfo{year}{2004}.
\newblock \bibinfo{title}{Methods for estimating the computational power and generalization capability of neural microcircuits}, in: \bibinfo{editor}{Saul, L.K.}, \bibinfo{editor}{Weiss, Y.}, \bibinfo{editor}{Bottou, L.} (Eds.), \bibinfo{booktitle}{Advances in Neural Information Processing Systems}, \bibinfo{publisher}{MIT Press}. pp. \bibinfo{pages}{865--872}.
\newblock \URLprefix \url{https://proceedings.neurips.cc/paper_files/paper/2004/file/9ff7c9eb9d37f434db778f59178012da-Paper.pdf}.
\bibitem[{Maass et~al.(2002)Maass, Natschl{\"a}ger and Markram}]{maass_lsm_2002}
\bibinfo{author}{Maass, W.}, \bibinfo{author}{Natschl{\"a}ger, T.}, \bibinfo{author}{Markram, H.}, \bibinfo{year}{2002}.
\newblock \bibinfo{title}{Real-time computing without stable states: A new framework for neural computation based on perturbations}.
\newblock \bibinfo{journal}{Neural Computation} \bibinfo{volume}{14}, \bibinfo{pages}{2531--2560}.
\newblock \DOIprefix\doi{10.1162/089976602760407955}.
\bibitem[{Matzner(2022)}]{Matzner2022HyperparameterTuningESN}
\bibinfo{author}{Matzner, F.}, \bibinfo{year}{2022}.
\newblock \bibinfo{title}{Hyperparameter tuning in echo state networks}, in: \bibinfo{booktitle}{Proceedings of the Genetic and Evolutionary Computation Conference}, \bibinfo{publisher}{ACM}, \bibinfo{address}{Boston, Massachusetts, USA}. pp. \bibinfo{pages}{404--412}.
\newblock \DOIprefix\doi{10.1145/3512290.3528721}.
\bibitem[{Morra and Daley(2022)}]{morra_daley_2022}
\bibinfo{author}{Morra, J.}, \bibinfo{author}{Daley, M.}, \bibinfo{year}{2022}.
\newblock \bibinfo{title}{Imposing connectome-derived topology on an echo state network}, in: \bibinfo{booktitle}{2022 International Joint Conference on Neural Networks (IJCNN)}, \bibinfo{publisher}{IEEE}. pp. \bibinfo{pages}{1--6}.
\newblock \DOIprefix\doi{10.1109/IJCNN55064.2022.9892629}.
\bibitem[{Morra et~al.(2023)Morra, Flynn, Amann and Daley}]{morra_fly_rc_2023}
\bibinfo{author}{Morra, J.}, \bibinfo{author}{Flynn, A.}, \bibinfo{author}{Amann, A.}, \bibinfo{author}{Daley, M.}, \bibinfo{year}{2023}.
\newblock \bibinfo{title}{Multifunctionality in a connectome-based reservoir computer}, in: \bibinfo{booktitle}{2023 IEEE International Conference on Systems, Man, and Cybernetics (SMC)}, \bibinfo{publisher}{IEEE}. pp. \bibinfo{pages}{4961--4966}.
\newblock \DOIprefix\doi{10.1109/SMC53992.2023.10394668}.
\bibitem[{Pilati et~al.(2026)Pilati, Ceni, Michieletti, Gallicchio, Ricciardi and Milano}]{pilati_ceni_michieletti_gallicchio_ricciardi_milano_2026}
\bibinfo{author}{Pilati, D.}, \bibinfo{author}{Ceni, A.}, \bibinfo{author}{Michieletti, F.}, \bibinfo{author}{Gallicchio, C.}, \bibinfo{author}{Ricciardi, C.}, \bibinfo{author}{Milano, G.}, \bibinfo{year}{2026}.
\newblock \bibinfo{title}{{RCbench}: A unified framework for benchmarking reservoir computing systems}.
\newblock \bibinfo{journal}{Neuromorphic Computing and Engineering} \bibinfo{volume}{6}, \bibinfo{pages}{014012}.
\newblock \DOIprefix\doi{10.1088/2634-4386/ae441f}.
\bibitem[{Platt et~al.(2022)Platt, Penny, Smith, Chen and Abarbanel}]{platt_systematic_2022}
\bibinfo{author}{Platt, J.A.}, \bibinfo{author}{Penny, S.G.}, \bibinfo{author}{Smith, T.A.}, \bibinfo{author}{Chen, T.C.}, \bibinfo{author}{Abarbanel, H.D.I.}, \bibinfo{year}{2022}.
\newblock \bibinfo{title}{A systematic exploration of reservoir computing for forecasting complex spatiotemporal dynamics}.
\newblock \bibinfo{journal}{Neural Networks} \bibinfo{volume}{153}, \bibinfo{pages}{530--552}.
\newblock \DOIprefix\doi{10.1016/j.neunet.2022.06.025}.
\bibitem[{Roy and Vetterli(2007)}]{roy_effective_rank_2007}
\bibinfo{author}{Roy, O.}, \bibinfo{author}{Vetterli, M.}, \bibinfo{year}{2007}.
\newblock \bibinfo{title}{The effective rank: A measure of effective dimensionality}, in: \bibinfo{booktitle}{Proceedings of the 15th European Signal Processing Conference (EUSIPCO 2007)}, \bibinfo{address}{Pozna{\'n}, Poland}. pp. \bibinfo{pages}{606--610}.
\newblock \URLprefix \url{https://eurasip.org/Proceedings/Eusipco/Eusipco2007/Papers/a5p-h05.pdf}.
\bibitem[{Salimans et~al.(2017)Salimans, Ho, Chen, Sidor and Sutskever}]{salimans_evolution_2017}
\bibinfo{author}{Salimans, T.}, \bibinfo{author}{Ho, J.}, \bibinfo{author}{Chen, X.}, \bibinfo{author}{Sidor, S.}, \bibinfo{author}{Sutskever, I.}, \bibinfo{year}{2017}.
\newblock \bibinfo{title}{Evolution strategies as a scalable alternative to reinforcement learning}.
\newblock \DOIprefix\doi{10.48550/arXiv.1703.03864}, \href{http://arxiv.org/abs/1703.03864}{\tt arXiv:1703.03864}.
\bibitem[{Srinivasan et~al.(2025)Srinivasan, Plenz and Girvan}]{srinivasan_ei_balance_2025}
\bibinfo{author}{Srinivasan, K.}, \bibinfo{author}{Plenz, D.}, \bibinfo{author}{Girvan, M.}, \bibinfo{year}{2025}.
\newblock \bibinfo{title}{Boosting reservoir computing with brain-inspired adaptive control of {E--I} balance}.
\newblock \bibinfo{journal}{Nature Communications} \bibinfo{volume}{16}, \bibinfo{pages}{10212}.
\newblock \DOIprefix\doi{10.1038/s41467-025-64978-8}.
\bibitem[{Stepney(2024)}]{Stepney2024PRCtutorial}
\bibinfo{author}{Stepney, S.}, \bibinfo{year}{2024}.
\newblock \bibinfo{title}{Physical reservoir computing: A tutorial}.
\newblock \bibinfo{journal}{Natural Computing} \bibinfo{volume}{23}, \bibinfo{pages}{665--685}.
\newblock \DOIprefix\doi{10.1007/s11047-024-09997-y}.
\bibitem[{Sukenik et~al.(2021)Sukenik, Vinogradov, Weinreb, Segal, Levina and Moses}]{sukenik_connection_numbers_2021}
\bibinfo{author}{Sukenik, N.}, \bibinfo{author}{Vinogradov, O.}, \bibinfo{author}{Weinreb, E.}, \bibinfo{author}{Segal, M.}, \bibinfo{author}{Levina, A.}, \bibinfo{author}{Moses, E.}, \bibinfo{year}{2021}.
\newblock \bibinfo{title}{Neuronal circuits overcome imbalance in excitation and inhibition by adjusting connection numbers}.
\newblock \bibinfo{journal}{Proceedings of the National Academy of Sciences of the United States of America} \bibinfo{volume}{118}, \bibinfo{pages}{e2018459118}.
\newblock \DOIprefix\doi{10.1073/pnas.2018459118}.
\bibitem[{Sun et~al.(2024)Sun, Song, Cai, Zhang, Hong and Li}]{sun_systematic_review_2024}
\bibinfo{author}{Sun, C.}, \bibinfo{author}{Song, M.}, \bibinfo{author}{Cai, D.}, \bibinfo{author}{Zhang, B.}, \bibinfo{author}{Hong, S.}, \bibinfo{author}{Li, H.}, \bibinfo{year}{2024}.
\newblock \bibinfo{title}{A systematic review of echo state networks from design to application}.
\newblock \bibinfo{journal}{IEEE Transactions on Artificial Intelligence} \bibinfo{volume}{5}, \bibinfo{pages}{23--37}.
\newblock \DOIprefix\doi{10.1109/TAI.2022.3225780}.
\bibitem[{Tanis et~al.(2009)Tanis, Bellemer, Moresco, Forbush and Koelle}]{tanis_bellemer_moresco_forbush_koelle_2009}
\bibinfo{author}{Tanis, J.E.}, \bibinfo{author}{Bellemer, A.}, \bibinfo{author}{Moresco, J.J.}, \bibinfo{author}{Forbush, B.}, \bibinfo{author}{Koelle, M.R.}, \bibinfo{year}{2009}.
\newblock \bibinfo{title}{The potassium chloride cotransporter {KCC-2} coordinates development of inhibitory neurotransmission and synapse structure in \emph{Caenorhabditis elegans}}.
\newblock \bibinfo{journal}{Journal of Neuroscience} \bibinfo{volume}{29}, \bibinfo{pages}{9943--9954}.
\newblock \DOIprefix\doi{10.1523/JNEUROSCI.1989-09.2009}.
\bibitem[{Varshney et~al.(2011)Varshney, Chen, Paniagua, Hall and Chklovskii}]{Varshney2011}
\bibinfo{author}{Varshney, L.R.}, \bibinfo{author}{Chen, B.L.}, \bibinfo{author}{Paniagua, E.}, \bibinfo{author}{Hall, D.H.}, \bibinfo{author}{Chklovskii, D.B.}, \bibinfo{year}{2011}.
\newblock \bibinfo{title}{Structural properties of the \emph{Caenorhabditis elegans} neuronal network}.
\newblock \bibinfo{journal}{PLOS Computational Biology} \bibinfo{volume}{7}, \bibinfo{pages}{e1001066}.
\newblock \DOIprefix\doi{10.1371/journal.pcbi.1001066}.
\bibitem[{Vidamour et~al.(2022)Vidamour, Ellis, Griffin, Venkat, Swindells, Dawidek, Broomhall, Steinke, Cooper, Maccherozzi, Dhesi, Stepney, Vasilaki, Allwood and Hayward}]{vidamour_quantifying_2022}
\bibinfo{author}{Vidamour, I.T.}, \bibinfo{author}{Ellis, M.O.A.}, \bibinfo{author}{Griffin, D.}, \bibinfo{author}{Venkat, G.}, \bibinfo{author}{Swindells, C.}, \bibinfo{author}{Dawidek, R.W.S.}, \bibinfo{author}{Broomhall, T.J.}, \bibinfo{author}{Steinke, N.J.}, \bibinfo{author}{Cooper, J.F.K.}, \bibinfo{author}{Maccherozzi, F.}, \bibinfo{author}{Dhesi, S.S.}, \bibinfo{author}{Stepney, S.}, \bibinfo{author}{Vasilaki, E.}, \bibinfo{author}{Allwood, D.A.}, \bibinfo{author}{Hayward, T.J.}, \bibinfo{year}{2022}.
\newblock \bibinfo{title}{Quantifying the computational capability of a nanomagnetic reservoir computing platform with emergent magnetisation dynamics}.
\newblock \bibinfo{journal}{Nanotechnology} \bibinfo{volume}{33}, \bibinfo{pages}{485203}.
\newblock \DOIprefix\doi{10.1088/1361-6528/ac87b5}.
\bibitem[{Watts and Strogatz(1998)}]{Wssmallworld}
\bibinfo{author}{Watts, D.J.}, \bibinfo{author}{Strogatz, S.H.}, \bibinfo{year}{1998}.
\newblock \bibinfo{title}{Collective dynamics of ``small-world'' networks}.
\newblock \bibinfo{journal}{Nature} \bibinfo{volume}{393}, \bibinfo{pages}{440--442}.
\newblock \DOIprefix\doi{10.1038/30918}.
\bibitem[{Wirak et~al.(2022)Wirak, Florman, Alkema, Connor and Gabel}]{wirak_florman_alkema_connor_gabel_2022}
\bibinfo{author}{Wirak, G.S.}, \bibinfo{author}{Florman, J.}, \bibinfo{author}{Alkema, M.J.}, \bibinfo{author}{Connor, C.W.}, \bibinfo{author}{Gabel, C.V.}, \bibinfo{year}{2022}.
\newblock \bibinfo{title}{Age-associated changes to neuronal dynamics involve a disruption of excitatory/inhibitory balance in \emph{C. elegans}}.
\newblock \bibinfo{journal}{eLife} \bibinfo{volume}{11}, \bibinfo{pages}{e72135}.
\newblock \DOIprefix\doi{10.7554/eLife.72135}.
\bibitem[{Yu et~al.(2026)Yu, Qin, Liu, Xu, Vogelstein, Brown and Vogelstein}]{yu_bpu_2025}
\bibinfo{author}{Yu, S.}, \bibinfo{author}{Qin, Z.}, \bibinfo{author}{Liu, T.}, \bibinfo{author}{Xu, B.}, \bibinfo{author}{Vogelstein, R.J.}, \bibinfo{author}{Brown, J.}, \bibinfo{author}{Vogelstein, J.T.}, \bibinfo{year}{2026}.
\newblock \bibinfo{title}{Biological processing units: Leveraging an insect connectome to pioneer biofidelic neural architectures}, in: \bibinfo{editor}{Ikl{\'e}, M.}, \bibinfo{editor}{Kolonin, A.}, \bibinfo{editor}{Bennett, M.} (Eds.), \bibinfo{booktitle}{Artificial General Intelligence: 18th International Conference, {AGI} 2025, Proceedings, Part II}, \bibinfo{publisher}{Springer}, \bibinfo{address}{Cham}. pp. \bibinfo{pages}{361--369}.
\newblock \DOIprefix\doi{10.1007/978-3-032-00800-8_32}. \bibinfo{note}{first published online August 7, 2025}.

\end{thebibliography}
\clearpage

\appendix
\section{Supplemental Material}

\subsection{Squared-correlation score}
\label{supp:r_score}
For target values \(y_t\) and predictions \(\hat{y}_t\) over evaluation samples indexed by \(t\), the score used for MC and the truncated IPC diagnostic is
\[
r_{\mathrm{score}}(y,\hat{y})
=
\operatorname{clip}_{[0,1]}\!\left[
\left(
\frac{\sum_t (y_t-\bar{y})(\hat{y}_t-\bar{\hat{y}})}
{\sqrt{\sum_t (y_t-\bar{y})^2\sum_t (\hat{y}_t-\bar{\hat{y}})^2}+\epsilon}
\right)^2
\right].
\]
Here, \(\bar{y}\) and \(\bar{\hat{y}}\) are the sample means of the target and predicted values, respectively; \(\epsilon=10^{-12}\) prevents division by zero; and \(\operatorname{clip}_{[0,1]}\) restricts numerical output to the interval from 0 to 1. Thus, \(r_{\mathrm{score}}\) is the squared Pearson correlation.

\subsection{Truncation sensitivity of MC and IPC}
\label{supp:metric_truncation}
We tested whether the choices \(D_{\max}=30\) and \(K=\{1,2,3,4,5\}\) materially affect comparisons among the ten architecture models. For these checks, \textit{capture} denotes the truncated capacity as a percentage of the corresponding value from an extended reference calculation, and ranking stability is calculated across architecture-level means.

For the polynomial-order check, the sum over orders 1--5 captured \(92.82\%\pm2.34\%\) of the value obtained using orders 1--10; the minimum capture was \(82.37\%\), and the fifth percentile was \(88.81\%\). Agreement between the two calculations was high (Pearson \(r=0.9993\), Spearman \(r_s=0.9979\)).
For the IPC delay check, values computed with \(D_{\max}=30\) captured \(91.10\%\)--\(96.33\%\) of the corresponding \(D_{\max}=80\) architecture means. For the empirical \textit{C.\ elegans} model, capture was \(92.16\%\pm1.22\%\). Despite the reduction in absolute IPC, the ranking of all ten architectures was unchanged relative to \(D_{\max}=80\) (Pearson \(r=1.0000\), Spearman \(r_s=1.0000\)).

MC was less sensitive to the delay cutoff. At \(D_{\max}=30\), the mean capture across architectures was \(97.86\%\), with architecture means ranging from \(96.90\%\) to \(98.89\%\). Capture for the empirical \textit{C.\ elegans} model was \(97.18\%\pm0.66\%\), and the architecture ranking was again identical to that obtained with \(D_{\max}=80\) (Pearson \(r=1.0000\), Spearman \(r_s=1.0000\)). Thus, the selected truncations modestly reduce absolute capacity values, particularly for IPC, but preserve the comparative architecture results.

\subsection{Common-tail sensitivity of GR}
\label{supp:gr_tail_sensitivity}

We tested whether the three-step common tail used for GR affects comparisons among the models shown in Figure~\ref{fig:arch}. The independent prefix was fixed at seven steps while the shared-tail length was varied over \(L_c=\{1,2,3,4,5,7,10\}\), with \(L_c=3\) as the reference.

Although GR decreased with longer shared tails, architecture ordering remained strongly preserved. Across positive tail lengths, correlations with the \(L_c=3\) architecture means were Pearson \(r=0.9835\)--\(1.0000\) and Spearman \(r_s=0.9000\)--\(1.0000\). GR values depend on the chosen tail length, while the relative differences among architectures are largely preserved across the tested tail lengths.

\subsection{Network perturbation procedures}
\label{supp:perturbation_procedures}
\subsubsection{Connection shuffle}
\label{supp:Connection-Shuffle}
We define a connection shuffle as a directed degree-preserving edge swap. Let \(W\) be the weighted adjacency matrix. We randomly select two existing directed edges \((a,b)\) and \((c,d)\), where \(a,b,c,d\) are distinct and the proposed edges \((a,d)\) and \((c,b)\) do not already exist. We then form a new matrix \(W'\) by setting
\[
W'_{ad}=W_{ab}, \qquad
W'_{cb}=W_{cd}, \qquad
W'_{ab}=0, \qquad
W'_{cd}=0,
\]
with all other entries unchanged. The procedure pairs edges without replacement and attempts up to \(\lfloor |E|/2\rfloor\) directed double-edge swaps. For the empirical \textit{C.\ elegans} network, \(|E|=3604\), giving a maximum of 1802 successful swaps per realization. In an audit of 5000 independent shuffle realizations, the number of successful swaps ranged from 1799 to 1802 (median 1801), corresponding to at least 99.83\% of the maximum in every realization. The shuffled networks retained, on average, 4.94\% of the original directed edges.

\subsubsection{Weight shuffle}

\label{supp:Weight-Shuffle}
Fix the synaptic connectivity pattern and randomly permute the weights assigned to existing connections. Let
\[
NZ=\{(i,j): W_{ij}\neq 0\}
\]
denote the set of nonzero entries in the weight matrix. We extract the corresponding weights into a vector,
\[
w_{NZ}=\{W_{ij}:(i,j)\in NZ\},
\]
shuffle this vector, and then reassign the shuffled weights back to the same nonzero indices. Thus, the set of existing connections is unchanged, but the weights assigned to those connections are redistributed across the fixed network topology.
\subsection{Architecture details}
\label{supp:architectures}
\begin{figure}[t]
    \centering
    \includegraphics[width=\linewidth]{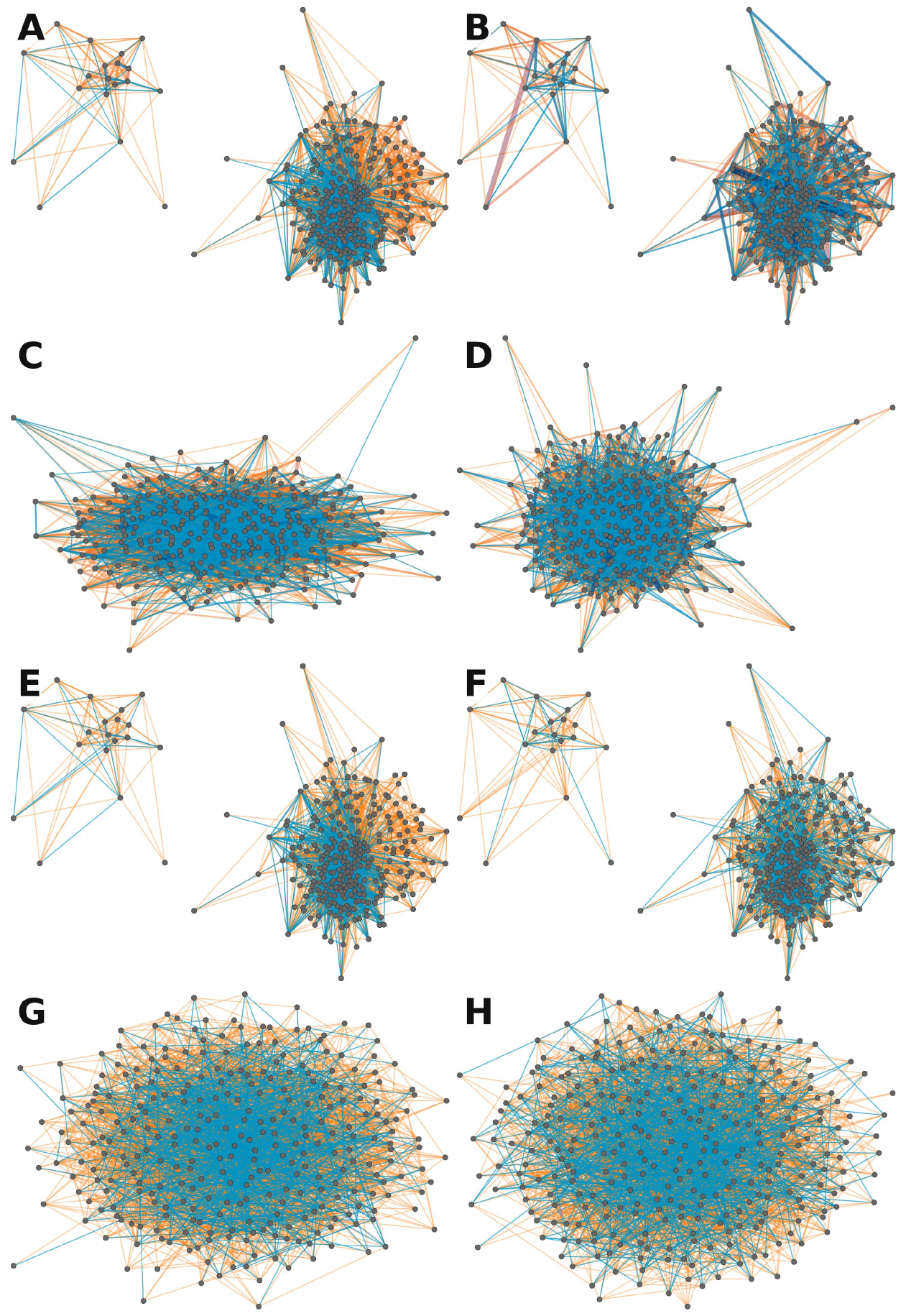}
    \caption[Network visualizations for architectures A--H]{Network visualizations for architectures A--H. In all panels, nodes represent neurons, orange edges are excitatory synapses, blue edges are inhibitory synapses, and darker edges indicate stronger synaptic weights. A: empirical \emph{C.~elegans} connectome. B: panel A with shuffled weights. C: panel A after connection shuffling. D: panel C with an additional weight shuffle. E: panel A topology with signed binary ($\pm 1$) weights determined by the empirical signs. F: panel E after sign shuffling. G: connection-shuffled version of panel E. H: panel G after sign shuffling.}
    \label{fig:a_to_h}
\end{figure}

\begin{figure}[t]
    \centering
    \includegraphics[width=\linewidth]{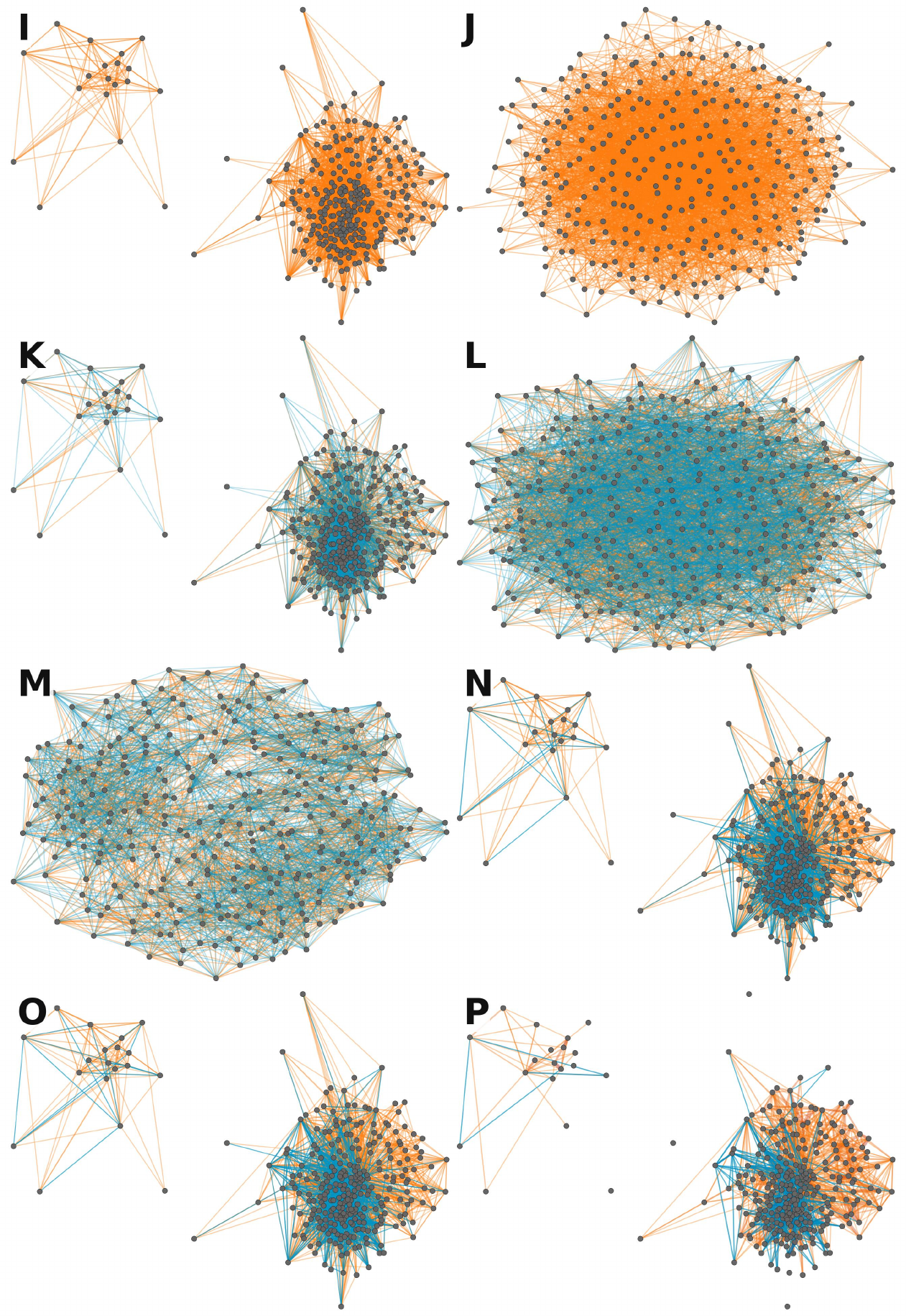}
    \caption[Additional control architectures I--P]{Additional control architectures I--P. I: empirical \emph{C.~elegans} topology with all-positive binary weights. J: panel I after connection shuffling. K: empirical \emph{C.~elegans} topology with Gaussian weights. L: \emph{C.~elegans}-matched Erd\H{o}s--R\'enyi network with Gaussian weights. M: small-world network with $p = 0.1$ and Gaussian weights. N: sign-preserving Gaussian model using the empirical signs. O: sign-preserving uniform model using the empirical signs. P: predicted-polarity-only \emph{C.~elegans} connectome after removing unknown or complex-sign chemical edges.}
    \label{fig:i_to_o}
\end{figure}

This section gives the full architecture definitions summarized in Table~\ref{tab:architecture_groups}. Figure~\ref{fig:a_to_h} shows panels A--H, which contain the shuffle controls. Figure~\ref{fig:i_to_o} shows panels I--P, which contain the additional binary, random-topology, and sign-preserving controls.

For direct comparison with the legends in Figures~\ref{fig:arch} and~\ref{fig:shuf_all}, each plotted architecture is marked below with its exact legend label and the figure or figures in which it appears. 

\subsubsection{C. elegans connectome}
\noindent\textbf{Label in Figures~\ref{fig:arch} and~\ref{fig:shuf_all}:} \enquote{C. elegans}.
This uses the biological network and weights:
\[
W_{ij}=W^{cel}_{ij}.
\]

\subsubsection{Empirical \textit{C.\ elegans} connectome with weight shuffle}
\label{sec:original_weight_shuffle}
\noindent\textbf{Figure~\ref{fig:shuf_all} label:} \enquote{shuffle}.
Starting from the empirical \textit{C.\ elegans} connectome, this architecture preserves the connectome topology but shuffles the weights across existing connections using the weight-shuffle procedure in Section~\ref{supp:Weight-Shuffle}.

\subsubsection{Empirical \textit{C.\ elegans} connectome with connection shuffle}
\label{sec:original_connection_shuffle}
\noindent\textbf{Figure~\ref{fig:shuf_all} label:} \enquote{Conn. shuffle}.
Starting from the empirical \textit{C.\ elegans} connectome, this architecture applies the degree-preserving connection-shuffle procedure in Section~\ref{supp:Connection-Shuffle}.

\subsubsection{Connection shuffle with weight shuffle}
\noindent\textbf{Figure~\ref{fig:shuf_all} label:} \enquote{Conn. + shuf.}.
The \textit{C.\ elegans} connectome is first connection-shuffled as described in Section~\ref{supp:Connection-Shuffle}, and the resulting nonzero weights are then shuffled as described in Section~\ref{supp:Weight-Shuffle}.

\subsubsection[C. elegans architecture with signed unit weights using empirical signs]{C. elegans architecture with $\pm 1$ weights using empirical signs}
\label{sec:plus1minus1network}
\noindent\textbf{Label in Figures~\ref{fig:arch} and~\ref{fig:shuf_all}:} \enquote{Sign-pres. pm1}.
This architecture preserves the empirical topology and edge signs while setting every nonzero magnitude to one:
\[
W_{ij}' =
\begin{cases}
\operatorname{sign}(W_{ij}) & \text{if } W_{ij} \neq 0,\\
0 & \text{if } W_{ij} = 0.
\end{cases}
\]

\subsubsection[C. elegans architecture with signed unit weights and shuffled signs preserving sign balance]{C. elegans architecture with $\pm 1$ weights and shuffled signs preserving sign balance}
\noindent\textbf{Label in Figures~\ref{fig:arch} and~\ref{fig:shuf_all}:} \enquote{pm1 + shuf.}.
This control keeps the empirical connectivity fixed but randomizes which edges carry inhibitory signs. If
\[
N_-=\#\{(i,j):W_{ij}<0\},
\]
then a random subset of size \(N_-\) of nonzero entries is assigned sign \(-1\), and all remaining nonzero entries are assigned sign \(+1\). This preserves the global E/I sign ratio while shuffling sign placement.

\subsubsection{Real weights magnitudes with shuffled signs preserving sign balance}
\label{supp:sign-pres_real_w}
\noindent\textbf{Figure~\ref{fig:arch} label:} \enquote{Sign shuf.}.
This control preserves the empirical topology, the magnitude on every edge, and the global number of negative edges, but randomizes which edges are negative. Let \(NZ=\{(i,j):W_{ij}\neq0\}\), let \(N_-\) be the number of negative entries in the empirical matrix, and select a subset \(S_-\subset NZ\) of size \(N_-\) uniformly at random. The control matrix is
\[
W'_{ij}=
\begin{cases}
s_{ij}|W_{ij}| & \text{if }(i,j)\in NZ,\\
0 & \text{otherwise},
\end{cases}
\]
where
\[
s_{ij}=
\begin{cases}
-1 & \text{if }(i,j)\in S_-,\\
+1 & \text{if }(i,j)\in NZ\setminus S_-.
\end{cases}
\]
Thus, ``Sign-pres.'' in this specific legend label refers to preservation of the global sign balance, not preservation of the original edge-by-edge sign placement.

\subsubsection[C. elegans architecture with signed unit weights under connection shuffling]{C. elegans architecture with $\pm 1$ weights under connection shuffling}
\noindent\textbf{Figure~\ref{fig:shuf_all} label:} \enquote{pm1 + conn. shuf.}.
Starting from the signed unit-magnitude network in Section~\ref{sec:plus1minus1network}, the edge set is connection-shuffled as described in Section~\ref{supp:Connection-Shuffle}.

\subsubsection[C. elegans architecture with signed unit weights under connection and weight shuffling]{C. elegans architecture with $\pm 1$ weights under connection shuffling and weight shuffling}
\noindent\textbf{Figure~\ref{fig:shuf_all} label:} \enquote{pm1 sign/conn/shuf.}.
Starting from the signed unit-magnitude network in Section~\ref{sec:plus1minus1network}, the edge set is connection-shuffled and then the signed unit weights are shuffled across the resulting nonzero entries.

\subsubsection{C. elegans architecture with binary weights}
\noindent\textbf{Label in Figures~\ref{fig:arch} and~\ref{fig:shuf_all}:} \enquote{Binary}.
This perturbation preserves the empirical connectivity pattern but removes both sign and magnitude heterogeneity:
\[
W_{ij}' =
\begin{cases}
1 & \text{if } W_{ij} \neq 0,\\
0 & \text{if } W_{ij} = 0.
\end{cases}
\]

\subsubsection{C. elegans architecture with binary weights and connection shuffle}
\noindent\textbf{Figure~\ref{fig:shuf_all} label:} \enquote{Binary shuf.}.
This binary network is rewired with the procedure in Section~\ref{supp:Connection-Shuffle}.

\subsubsection{C.~elegans topology with Gaussian weights}
\noindent\textbf{Figure~\ref{fig:arch} label:} \enquote{CE Gaussian}.
This control keeps the empirical connectivity pattern but replaces each nonzero weight with an independent Gaussian sample:
\[
W'_{ij} =
\begin{cases}
X_{ij} & \text{if } W_{ij} \neq 0,\\
0 & \text{if } W_{ij} = 0,
\end{cases}
\qquad
X_{ij}\sim\mathcal{N}(0,1).
\]

\subsubsection{C.~elegans--matched Erd\H{o}s--R\'enyi model with Gaussian weights}
\noindent\textbf{Figure~\ref{fig:arch} label:} \enquote{ER Gaussian}.
This random-graph baseline uses \(N=297\) nodes and an adjusted ER edge set with the same number of directed edges as the empirical \emph{C.~elegans} network. After fixing the edge count, weights are independently sampled from \(\mathcal{N}(0,1)\).

\subsubsection{Watts--Strogatz Gaussian control}
\noindent\textbf{Figure~\ref{fig:arch} label:} \enquote{WS Gaussian}.
This small-world control uses \(N=297\) nodes and \(3,604\) directed edges. A directed ring lattice is rewired with probability \(p=0.1\), preserving source out-degree and edge count, and nonzero weights are independently sampled from \(\mathcal{N}(0,1)\).

\subsubsection{C. elegans architecture with half Gaussian weights using empirical signs}
\noindent\textbf{Figure~\ref{fig:arch} label:} \enquote{Sign-pres. Gaussian}.
Starting from the signed unit-magnitude network in Section~\ref{sec:plus1minus1network}, magnitudes are sampled from a half-Gaussian distribution:
\[
W_{ij}'=\operatorname{sign}(W_{ij})|\mathcal{N}(0,1)|.
\]

\subsubsection{C. elegans architecture with uniform weights using empirical signs}
\noindent\textbf{Figure~\ref{fig:arch} label:} \enquote{Sign-pres. uniform}.
Starting from the signed unit-magnitude network in Section~\ref{sec:plus1minus1network}, magnitudes are sampled independently from a uniform distribution:
\[
W_{ij}'=\operatorname{sign}(W_{ij})X_{ij},
\qquad
X_{ij}\sim\mathcal{U}(0,1).
\]

\subsection{Additional E/I edge balance sweep controls}
\label{supp:sign_fraction_controls}
\begin{figure}[p]
    \centering
    \includegraphics[width=\linewidth]{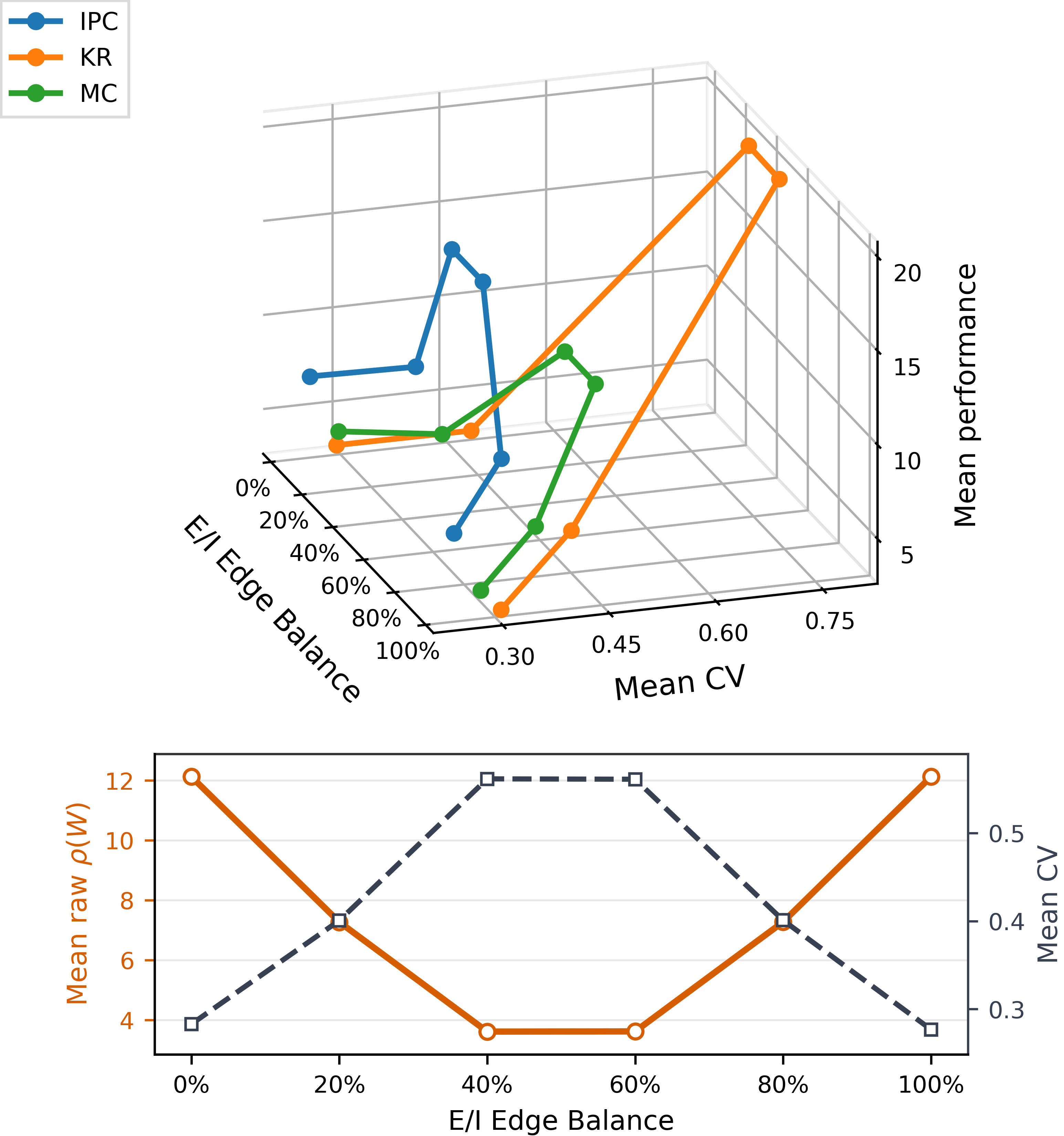}
    \caption{E/I edge balance sweep for the \textit{C.\ elegans}-matched ER topology. Panel A shows mean task-agnostic performance and mean hyperparameter CV as a function of E/I edge balance. The truncated IPC diagnostic is denoted in orange, MC in blue, and KR in green. The same performance--robustness tradeoff appears without the biological topology. Panel B shows the mean raw spectral radius, \(\rho_{\mathrm{raw}}\), before normalization (solid orange curve with circular markers; left axis) and the mean CV averaged across the three task-agnostic metrics (black dashed curve with square markers; right axis) across the same E/I sweep. Points show averages across 200 trials for each fraction.}
    \label{fig:er_frac_curves}
\end{figure}

\begin{figure}[p]
    \centering
    \includegraphics[width=\linewidth]{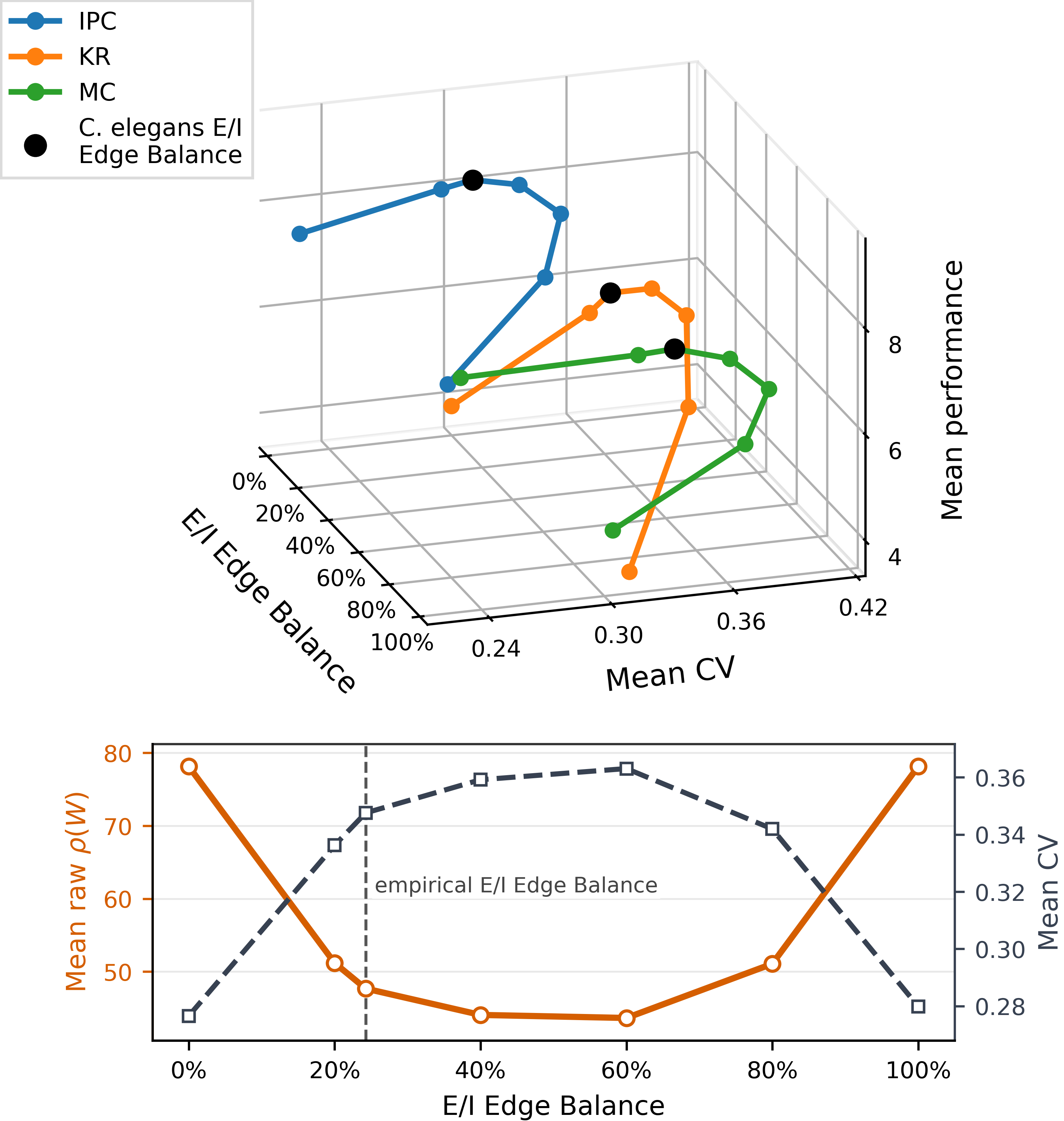}
    \caption{E/I edge balance sweep for the predicted-polarity-only \textit{C.\ elegans} magnitude pattern after removing connections with unknown or complex signs. Panel A shows mean task-agnostic performance and mean hyperparameter CV as a function of E/I edge balance. The truncated IPC diagnostic is denoted in orange, MC in blue, and KR in green; The three black markers in Panel A identify the values of the three task-agnostic metrics at the empirical \textit{C.\ elegans} E/I edge balance. Panel B shows the mean raw spectral radius, \(\rho_{\mathrm{raw}}\), before normalization (solid orange curve with circular markers; left axis) and the mean CV averaged across the three task-agnostic metrics (black dashed curve with square markers; right axis) across the same E/I sweep. The vertical gray dashed line marks the empirical \textit{C.\ elegans} E/I edge balance of approximately \(24.25\%\). Points show averages across 200 trials for each fraction.}
    \label{fig:removed_frac}
\end{figure}

The main text shows the E/I edge balance sweep for the empirical \textit{C.\ elegans} graph in Figure~\ref{fig:ambig_cel_frac_curves}. The matched ER control is shown in Figure~\ref{fig:er_frac_curves}, and the predicted-polarity-only \textit{C.\ elegans} control is shown in Figure~\ref{fig:removed_frac}. In each figure, Panel A shows the performance--robustness tradeoff and Panel B shows the associated mean raw spectral radius before normalization. Together, these comparisons show that both trends appear across multiple topology and preprocessing conditions under spectral-radius normalization.

\subsection{Shuffle experiment CV summaries}
\label{supp:shuffle_cv}
Tables~\ref{tab:shuffle_cv_real}--\ref{tab:shuffle_cv_binary} report the paired changes in mean hyperparameter CV for the shuffle experiments shown in Figure~\ref{fig:shuf_all}. Positive values indicate higher CV than the corresponding baseline, and negative values indicate lower CV. Each comparison uses \(n=200\) paired trial groups.

\begin{table}[p]
    \centering
    \scriptsize
    \caption[Mean CV differences for real-valued shuffle controls]{Mean CV differences for real-valued shuffle controls relative to the empirical \textit{C.\ elegans} reservoir. Baseline CV and control CV are mean coefficients of variation across the 96-point hyperparameter sweep. \(\Delta\)CV is control minus baseline.}
    \label{tab:shuffle_cv_real}
    \resizebox{\columnwidth}{!}{\begin{tabular}{llrrrrl}
\toprule
Metric & Control & C. elegans mean CV & Control mean CV & $\Delta$CV & $\Delta$CV (\%) & 95\% CI \\
\midrule
MC & Conn. + wt. shuf. & 0.4088 & 0.4983 & +0.0895 & +21.89 & [+0.0837, +0.0953] \\
MC & Conn. shuffle & 0.4088 & 0.4974 & +0.0886 & +21.68 & [+0.0831, +0.0942] \\
MC & Wt. shuffle & 0.4088 & 0.4895 & +0.0807 & +19.73 & [+0.0749, +0.0865] \\
IPC & Conn. + wt. shuf. & 0.2995 & 0.4049 & +0.1054 & +35.18 & [+0.1011, +0.1096] \\
IPC & Conn. shuffle & 0.2995 & 0.3948 & +0.0953 & +31.81 & [+0.0911, +0.0995] \\
IPC & Wt. shuffle & 0.2995 & 0.3935 & +0.0940 & +31.38 & [+0.0892, +0.0988] \\
KR & Conn. + wt. shuf. & 0.3980 & 0.5920 & +0.1940 & +48.74 & [+0.1882, +0.1998] \\
KR & Conn. shuffle & 0.3980 & 0.6069 & +0.2090 & +52.51 & [+0.2029, +0.2151] \\
KR & Wt. shuffle & 0.3980 & 0.5700 & +0.1720 & +43.23 & [+0.1660, +0.1781] \\
GR & Conn. + wt. shuf. & 0.3470 & 0.6444 & +0.2974 & +85.70 & [+0.2893, +0.3054] \\
GR & Conn. shuffle & 0.3470 & 0.6596 & +0.3126 & +90.09 & [+0.3034, +0.3218] \\
GR & Wt. shuffle & 0.3470 & 0.6246 & +0.2776 & +80.00 & [+0.2697, +0.2855] \\
\bottomrule
\end{tabular}
}
\end{table}

\begin{table}[p]
    \centering
    \scriptsize
    \caption[Mean CV differences for signed-unit shuffle controls]{Mean CV differences for signed-unit shuffle controls relative to the empirical local sign-preserving signed-unit reservoir. Baseline CV and control CV are mean coefficients of variation across the 96-point hyperparameter sweep. \(\Delta\)CV is control minus baseline.}
    \label{tab:shuffle_cv_pm1}
    \resizebox{\columnwidth}{!}{\begin{tabular}{llrrrrl}
\toprule
Metric & Control & Sign-pres. pm1 mean CV & Control mean CV & $\Delta$CV & $\Delta$CV (\%) & 95\% CI \\
\midrule
MC & pm1 + conn. shuf. & 0.4526 & 0.4202 & -0.0324 & -7.15 & [-0.0362, -0.0286] \\
MC & pm1 + wt. shuf. & 0.4526 & 0.3950 & -0.0576 & -12.73 & [-0.0612, -0.0540] \\
MC & pm1 sign/conn/wt shuf. & 0.4526 & 0.3909 & -0.0617 & -13.63 & [-0.0652, -0.0581] \\
IPC & pm1 + conn. shuf. & 0.3797 & 0.3574 & -0.0223 & -5.87 & [-0.0266, -0.0180] \\
IPC & pm1 + wt. shuf. & 0.3797 & 0.3179 & -0.0618 & -16.29 & [-0.0656, -0.0580] \\
IPC & pm1 sign/conn/wt shuf. & 0.3797 & 0.3197 & -0.0600 & -15.80 & [-0.0635, -0.0565] \\
KR & pm1 + conn. shuf. & 0.5386 & 0.4757 & -0.0629 & -11.68 & [-0.0701, -0.0557] \\
KR & pm1 + wt. shuf. & 0.5386 & 0.4263 & -0.1122 & -20.84 & [-0.1191, -0.1054] \\
KR & pm1 sign/conn/wt shuf. & 0.5386 & 0.4254 & -0.1132 & -21.02 & [-0.1200, -0.1064] \\
GR & pm1 + conn. shuf. & 0.5207 & 0.4833 & -0.0374 & -7.19 & [-0.0482, -0.0267] \\
GR & pm1 + wt. shuf. & 0.5207 & 0.4159 & -0.1048 & -20.13 & [-0.1148, -0.0948] \\
GR & pm1 sign/conn/wt shuf. & 0.5207 & 0.4246 & -0.0961 & -18.46 & [-0.1062, -0.0860] \\
\bottomrule
\end{tabular}
}
\end{table}

\begin{table}[p]
    \centering
    \scriptsize
    \caption[Mean CV differences for the all-positive binary topology shuffle]{Mean CV differences for the all-positive binary topology shuffle relative to the all-positive binary \textit{C.\ elegans} topology. Baseline CV and control CV are mean coefficients of variation across the 96-point hyperparameter sweep. \(\Delta\)CV is control minus baseline.}
    \label{tab:shuffle_cv_binary}
    \resizebox{\columnwidth}{!}{\begin{tabular}{llrrrrl}
\toprule
Metric & Control & Binary wt. mean CV & Control mean CV & $\Delta$CV & $\Delta$CV (\%) & 95\% CI \\
\midrule
MC & Binary shuf. & 0.2982 & 0.2678 & -0.0305 & -10.21 & [-0.0332, -0.0278] \\
IPC & Binary shuf. & 0.2317 & 0.2302 & -0.0015 & -0.66 & [-0.0044, +0.0013] \\
KR & Binary shuf. & 0.2735 & 0.2622 & -0.0113 & -4.13 & [-0.0134, -0.0092] \\
GR & Binary shuf. & 0.2192 & 0.2347 & +0.0155 & +7.07 & [+0.0112, +0.0198] \\
\bottomrule
\end{tabular}
}
\end{table}


\subsection{Reservoir activity and closed-triad weights}
\begin{figure}[!b]
    \centering
    \includegraphics[width=\linewidth]{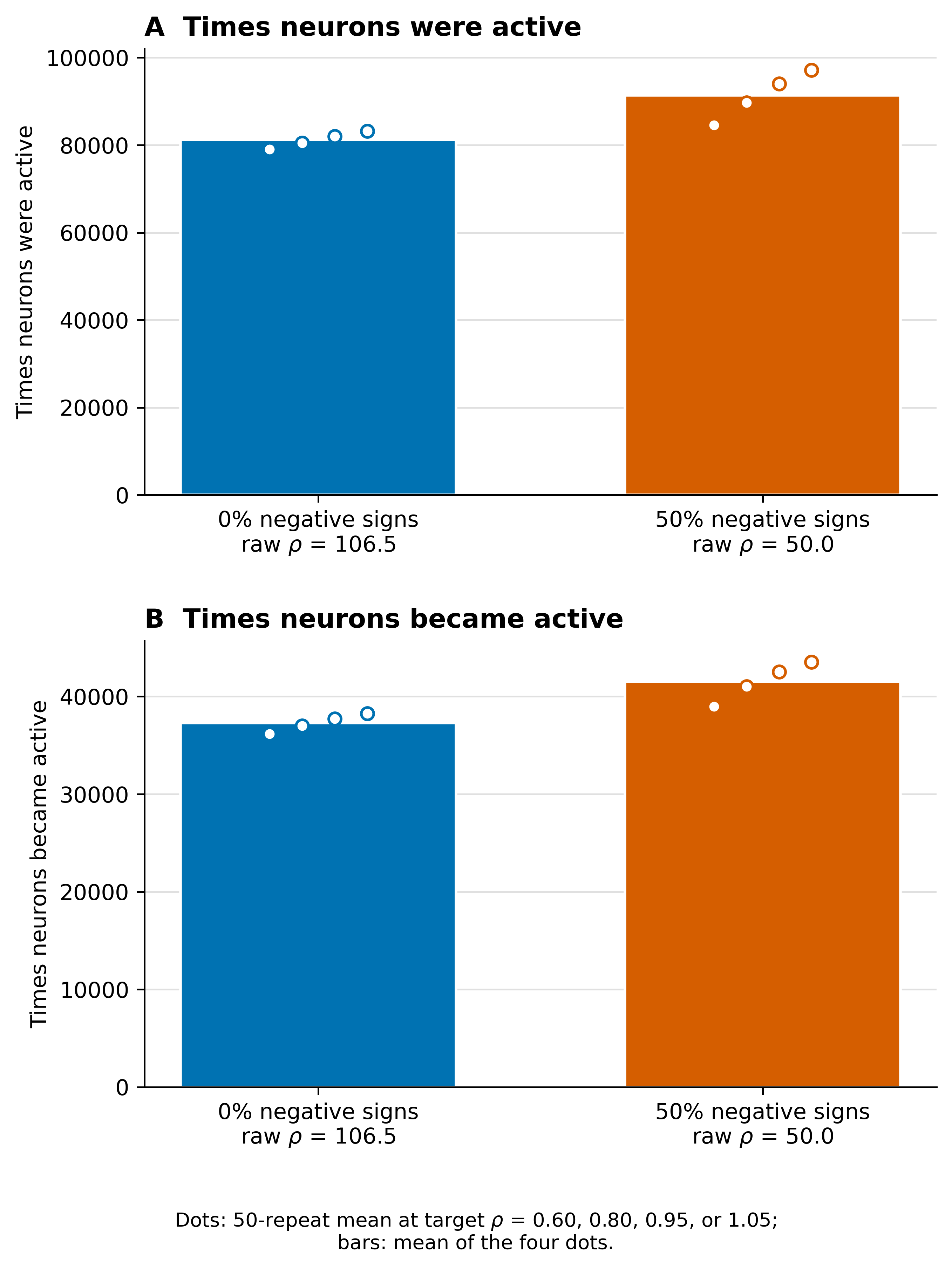}
    \caption{Reservoir activity in all-positive (0\% negative edges) and sign-balanced (50\% negative edges) versions of the empirical \textit{C.\ elegans} network. We count a neuron as active when \( |x_i|\geq0.5 \). Panel A counts how often neurons are active during 1,000 time steps. Panel B counts how often neurons change from inactive to active. The network connections and weight sizes are the same in both versions, but the signs are reassigned. Each dot is the mean across 50 repeats at one target spectral radius, \(\rho_{\mathrm{target}}\in\{0.60,0.80,0.95,1.05\}\); each bar is the mean of its four dots. The displayed raw spectral radii are means calculated before normalization.}
    \label{fig:activity_0_vs_50}
\end{figure}

\begin{figure}[!t]
    \centering
    \includegraphics[width=\linewidth]{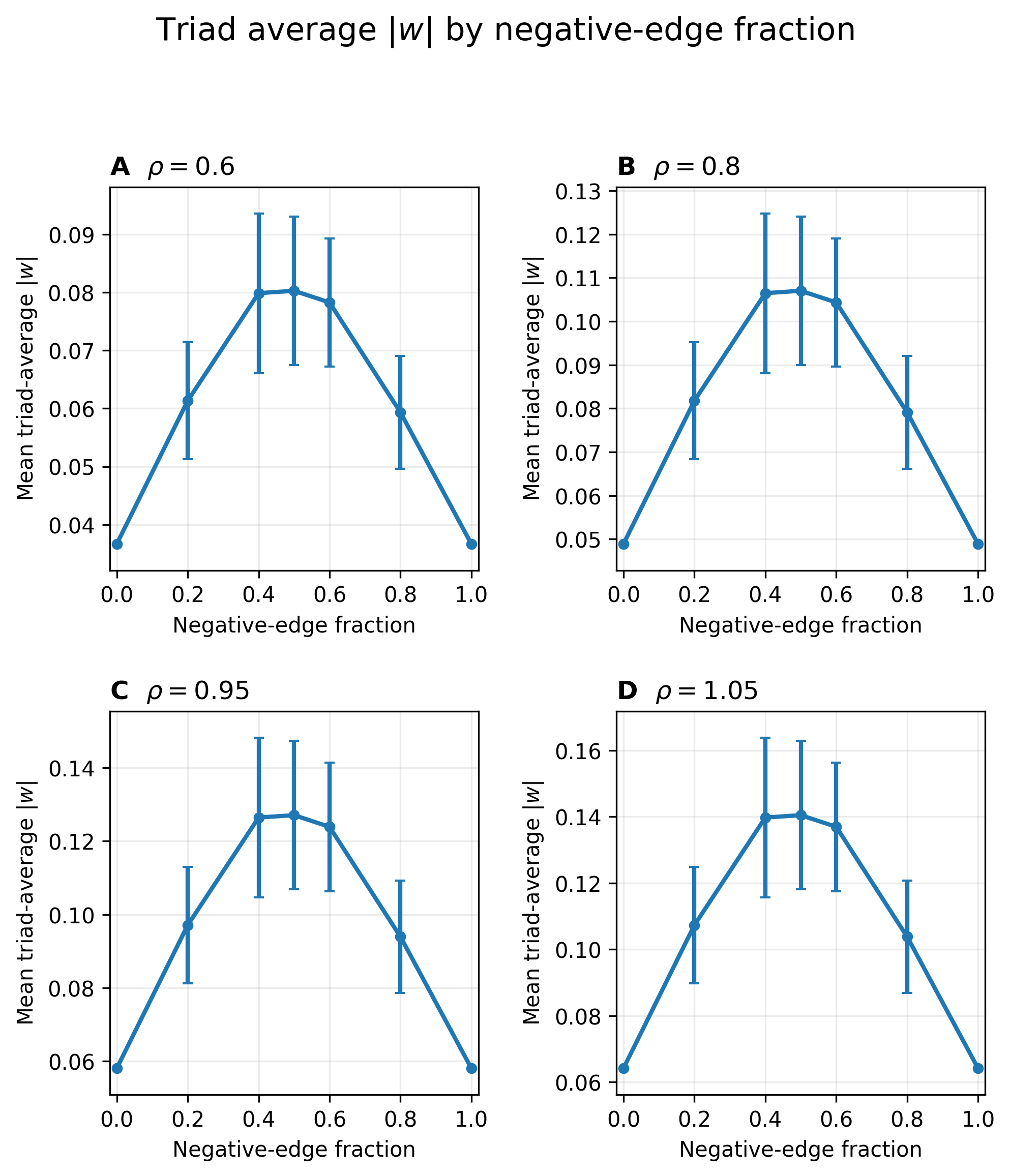}
    \caption{Mean absolute recurrent weight within closed three-neuron subgraphs across the E/I edge balance sweep. A closed triad is defined here as a set of three neurons for which every unordered neuron pair is connected by at least one directed edge; each triad can therefore contain three to six directed edges. For each of the 6,527 closed triads, we calculate the mean absolute weight over its present directed edges after scaling the complete reservoir to the target spectral radius. Points show the mean of this triad-level statistic across 50 repeats, and error bars show the standard deviation across repeat-level means. Panels correspond to target spectral radii \(\rho_{\mathrm{target}}=0.60\), \(0.80\), \(0.95\), and \(1.05\). Because topology and unnormalized weight magnitudes are held fixed, the variation across E/I edge balance fractions reflects changes in the global scaling required to reach each target spectral radius.}
    \label{fig:triad_sign_fraction}
\end{figure}

These analyses show how activity and recurrent weights change with sign balance and normalization. Across four target radii, the sign-balanced reservoir is active more often, and its neurons transition into an active state more often, than those in the all-positive reservoir (Figure~\ref{fig:activity_0_vs_50}). Across the full sweep, normalized weights in closed triads are smallest near the endpoints and largest near an even balance of positive and negative signs (Figure~\ref{fig:triad_sign_fraction}). As the raw spectral radius falls toward this balance (Figure~\ref{fig:ambig_cel_frac_curves}B), more scaling is needed to reach the target radius. Thus, the balanced networks combine stronger normalized triad weights with greater activity, although normalization contributes to both effects.

\clearpage
\addtocounter{page}{-1}
\end{document}